\documentclass[twocolumn]{biophys}
\usepackage{helvet,times}
\usepackage{bm,textcomp}

\jno{kxl014} %journal number
\gridframe{N}%option for grid around the text "Y" or "N"
\cropmark{N}%option for cropmark around the text "Y" or "N"

\usepackage{amssymb,amsmath,amsbsy,wasysym}
\usepackage[USenglish]{babel}
\usepackage{hyphenat}
\usepackage[hyperref]{xcolor}
\usepackage{bm}
\usepackage{subfigure}
\usepackage{enumerate}
\usepackage{color}

\usepackage{soul} % use this (many fancier options)

\usepackage{xr}
\usepackage[breaklinks=true,
            pdfborder={0 0 1},
            colorlinks=true,
            linkcolor=black,
            citecolor=blue,
            urlcolor=blue]{hyperref}
\usepackage[round,numbers,sort&compress]{natbib}

\newcommand{\ead}{e_\text{ad}}

\newcommand{\kt}{k_\text{B}T}

\newcommand{\rdomain}{r_\text{domain}}
\newcommand{\tinject}{\tau_\text{inject}}
\newcommand{\ltension}{\gamma}
\begin{document}

\setcounter{page}{1} %first page number

\title{Virus Assembly on a Membrane is Facilitated by Membrane Microdomains}

\author{Teresa Ruiz-Herrero $^{\dagger \footnote{Current affiliation: School of Engineering and Applied Sciences, Harvard University, Cambridge, MA }}$ , and Michael F. Hagan $^\ddagger$}

\address{$^\dagger$  Departamento de F\'isica Te\'orica de la Materia Condensada, Universidad Aut\'onoma de Madrid, Madrid, Spain;\\
$^\ddagger$ Martin Fisher School of Physics, Brandeis University, Waltham, Massachusetts}

% generate the title page from the info in the headers above

%Abstract environment needs 3 arguments. They are
%1. The abstract
%2. Received date
%3. Address, email

\begin{abstract}%
{For many viruses assembly and budding occur simultaneously during virion formation. Understanding the mechanisms underlying this process could promote biomedical efforts to block viral propagation and enable use of capsids in nanomaterials applications. To this end, we have performed molecular dynamics simulations on a coarse-grained model that describes virus assembly on a fluctuating lipid membrane. Our simulations show that the membrane can promote association of adsorbed subunits through dimensional reduction, but also can introduce barriers that inhibit complete assembly. We find several mechanisms, including one not anticipated by equilibrium theories, by which membrane microdomains, such as lipid rafts, can enhance assembly by reducing these barriers. We show how this predicted mechanism can be experimentally tested. Furthermore, the simulations demonstrate that assembly and budding depend crucially on the system dynamics via multiple timescales related to membrane deformation, protein diffusion, association, and adsorption onto the membrane.
}%1
{Insert Received for publication Date and in final form Date.}%2
{Correspondence: hagan@brandeis.edu}%3
\end{abstract}

\maketitle %%The above information typeset through this command

\section*{Introduction}

Processes in which proteins assemble on membranes to drive topology changes are ubiquitous in biology.  Despite extensive experimental and theoretical investigations (e.g. \cite{Baumgart2011,Krauss2011}), how assembly-driven membrane deformation depends on protein properties, membrane properties, and membrane compositional inhomogeneity remains incompletely understood. An important example of this phenomenon occurs during the formation of an enveloped virus, when the virion acquires a membrane envelope by budding from its host cell. Budding is typically driven at least in part by assembly of capsid proteins or viral membrane proteins \cite{Sundquist2012,Hurley2010,Welsch2007,Solon2005,Vennema1996,Garoff1998},
and many enveloped viruses, including HIV and influenza, preferentially bud from membrane microdomains (e.g. lipid rafts) \cite{Waheed2010,Welsch2007,Rossman2011}. Understanding how viruses exploit membrane domain structures to facilitate budding
would reveal fundamental aspects of the viral lifecycle, and could focus efforts to identify targets for new antiviral drugs that interfere with budding.  Furthermore, there is much interest in developing enveloped viral nanoparticles as targeted transport vehicles equipped to cross cell membranes through fusion \cite{Lundstrom2009,Cheng2013,Rowan2010}.  More generally, identifying the factors that make viral budding robust will shed light on other biological processes in which high-order complexes assemble to reshape membranes. Toward this goal, we perform dynamical simulations in which capsids simultaneously assemble and bud from model lipid membranes. We identify mechanisms by which membrane adsorption either promotes or impedes assembly, and we find multiple mechanisms by which a membrane microdomain significantly enhances assembly and budding.

Enveloped viruses can be divided into two groups based on how they acquire their lipid membrane envelope.  For the first group, which includes influenza and type C retroviruses (e.g. HIV), the (immature) nucleocapsid core assembles on the membrane concomitant with budding. In the second group, a core assembles in the cytoplasm prior to envelopment (reviewed in \cite{Sundquist2012,Welsch2007,Hurley2010}). In many families from this group (e.g. alphavirus) envelopment is driven by assembly of viral transmembrane glycoproteins around the core \cite{Garoff2004}.  For all enveloped viruses, membrane deformation is driven at least in part by a combination of weak protein-protein and protein-lipid interactions.  Thus, properties of the membrane should substantially affect budding and assembly timescales. In support of this hypothesis, many viruses from both groups preferentially bud from membrane microdomains 10-100 nm in size that are concentrated with cholesterol and/or sphingolipids \cite{Waheed2010, Welsch2007,
Rossman2011}. A critical question is whether viruses utilize microdomains primarily to concentrate capsid proteins or other molecules, or if the geometric and physical properties of domains facilitate budding. Answering these questions through experiments alone has been challenging \cite{Sundquist2012,Welsch2007,Hurley2010}.

 Extensive previous theoretical investigations have studied  budding by pre-assembled cores or nanoparticles (e.g. \cite{Ruiz-Herrero2012,Chaudhuri2011,Deserno2002,Fosnaric2009,Ginzburg2007,Jiang2008,Li2010a, Li2010,Smith2007,Tzlil2004,Vacha2011,Yang2011}), budding triggered by non-assembling subunits \cite{Reynwar2007}, or used a continuum model to study assembly and budding  \cite{Zhang2008}. Most closely related to our work, Matthews and Likos recently performed simulations on a coarse-grained model of patchy colloidal particles assembling on a membrane represented as a triangulated surface \cite{Matthews2012,Matthews2013,Matthews2013a}.  These elegant simulations provided a first look at the process of simultaneous assembly and budding and showed that subunit adsorption onto a membrane facilitates assembly through dimensional reduction. Here, we perform dynamical simulations on a model which more closely captures the essential geometric features of capsid subunits and lipid bilayers, and we explore how the presence of a microdomain within the membrane can influence assembly and budding. Our simulations show that, while the membrane can promote assembly of partial capsids, membrane deformations can introduce barriers that hinder completion of assembly. We find that a microdomain within a certain size range favors membrane deformations that diminish these barriers, and thus can play a key role in enabling complete assembly and budding. Furthermore, our simulations suggest that assembly morphologies depend crucially on multiple timescales, including those of protein-protein association, membrane deformation, and protein adsorption onto the membrane. Finally, we discuss potential effects of simplifications in our coarse-grained model and how a key prediction from the simulations can be tested in an {\it in vitro} assay.

 \begin{figure}
\centerline{ \includegraphics[viewport=0 0 1193 665, width=0.48\textwidth]{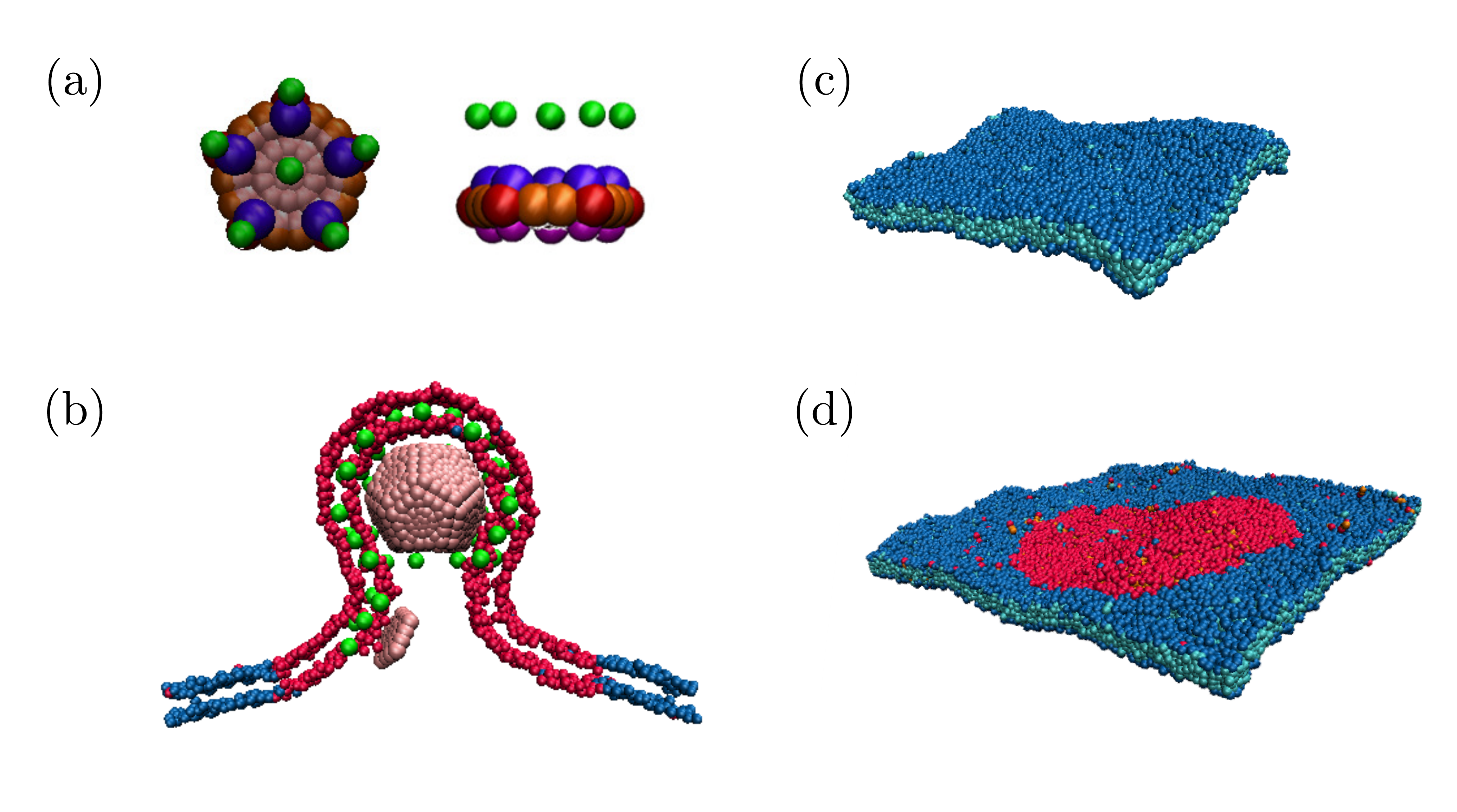}}
% model.pdf: 1193x665 pixel, 72dpi, 42.09x23.46 cm, bb=0 0 1193 665
\caption{The capsomer and membrane models. {\bf (a)} Top and side view of the capsomer. Attractive sites are red and orange, top and bottom repulsive sites are violet and magenta, excluders are pink, and capsomer-lipid interaction sites are green, with the pseudoatom types defined in Methods and in Supplementary Section ~\ref{sec:capsomer_model}. {\bf (b)} A slice of the membrane and the entire capsid are shown during budding, with the capsomer-lipid interaction sites colored green, and the domain lipids colored purple. {\bf (c)} A homogeneous membrane patch, with blue and cyan beads representing the lipid heads and lipid tails respectively.  {\bf (d)} A two-phase membrane, with red  and orange beads representing the domain lipid heads and tails respectively.  Images were generated using VMD \cite{Humphrey1996}.}
\label{fig:model}
\end{figure}

\section*{Methods}
Due to the large length and time scales associated with assembly of a capsid, simulating the process with an all-atom model is well beyond the capabilities of current computers \cite{Freddolino2006}. Therefore, in this paper we aim to elucidate the principles underlying simultaneous assembly and budding by considering a simplified geometric model for capsid proteins, inspired by previous simulations of empty capsid assembly \cite{Schwartz1998,Hagan2006,Hicks2006,Nguyen2007,Wilber2007,Nguyen2008,Nguyen2009,Johnston2010,Wilber2009a,Wilber2009,Rapaport1999,Rapaport2004,Rapaport2008,Hagan2011,Ayton2010,Chen2011} and assembly around nucleic acids \cite{Perlmutter2013, Elrad2010,Hagan2011,Mahalik2012,Zhang2013}.  Similarly, we consider a simplified model for lipids \cite{Cooke2005b, Reynwar2007} which recapitulates the material properties of biological membranes.

\subsection*{Membrane Model}

The membrane is represented by the model from Cooke and Deserno \cite{Cooke2005b}, in which each amphiphile is represented by one head bead and two tail beads connected by FENE bonds (Fig.~\ref{fig:model}c). This is an implicit solvent model; hydrophobic forces responsible for the formation of bilayers are mimicked by attractive interactions between tail beads with interaction strength $\epsilon_{0}$. This model enables computational feasibility while allowing the formation of bilayers with physical properties such as fluidity, diffusivity, and rigidity that are easily tuned across the range of values measured in biological membranes \cite{Cooke2005b,Ruiz-Herrero2012}. The bead diameter is set to $\sigma=0.9$ nm to obtain bilayers with widths of 5 nm and the lipid-lipid interaction strength is set to $\epsilon_0=1.1\kt$ to obtain fluid membranes with bending modulus $\kappa=8.25\kt$. When studying the effect of a domain, we consider two types of lipids, with \textit{M} and \textit{D} referring respectively to the lipids outside and inside of the domain, and  tail-tail interaction parameters $\epsilon_{ij}$ ( eq.~\ref{eq:hydrophobicity} in Supporting Information)
 set to $\epsilon_\text{DD}=\epsilon_\text{MM}=\epsilon_{0}$, while $\epsilon_\text{DM}$ is a variable parameter that controls the line tension of the domain, $\ltension$. Varying $\epsilon_\text{DM}$ from 0 to $\epsilon_0$ tunes the line tension from $\ltension_\text{max}$ to 0 (SI Sec.~\ref{sec:raft_model}). Further details for the membrane model are provided in SI Sec.  ~\ref{sec:raft_model}.

\subsection*{Capsid subunit model}
We modified and extended a model for assembly of non-enveloped capsids \cite{Wales2005, Fejer2009, Johnston2010,Perlmutter2013} to describe assembly on a membrane. A complete listing of the interaction potentials is provided in SI Section ~\ref{sec:capsomer_model};
we summarize them here. The capsid subunit is a rigid body with a pentagonal base and radius of $r_\text{pentamer}=5 \sigma$ formed by 15 attractive and 10 repulsive interaction sites (Fig.  \ref{fig:model} a). Subunit assembly is mediated through a Morse potential between `attractor' pseudoatoms located in the pentagon plane, with one located at each subunit vertex and 2 along each edge. Attractions occur between like attractors only, meaning that there are vertex-vertex and edge-edge attractions, but no vertex-edge attractor interactions. The 10 repulsive interaction sites are arranged symmetrically above and below the pentagon plane, so as to favor a subunit-subunit angle consistent with a dodecahedron (116 degrees). The motivation for our modifications to the model are described in SI Section \ref{sec:capsomer_model}.

\subsection*{Membrane-capsomer interaction}
The potential between capsomers and lipids accounts for attractive interactions and excluded-volume. First, we add to the capsomer body six attractor pseudoatoms that have attractive interactions with lipid tail beads. When simulating a phase-separated membrane, the attractors interact only with the domain lipid tails (Fig. \ref{fig:model}b). The attractors are placed one above each vertex and one above the center of the pentagon, each located a distance of $6\sigma$ above the pentagon plane (Fig. \ref{fig:model}a). These are motivated by, e.g., the myristate group on retrovirus GAG proteins that promotes subunit adsorption by inserting into the lipid bilayer \cite{Hamard-Peron2011}. The attractor-tail interaction is the same form as the lipid tail-tail  interaction except that there is no repulsive component (SI eq. (\ref{eq:adh_hydrophobicity})).  It is parameterized by the interaction strength,  $\epsilon_\text{ad}$, which tunes the adhesion energy per capsomer according to $e_\text{ad}=\alpha\epsilon_\text{ad}$ with $\alpha=-135.3\kt$ (SI Sec. \ref{sec:adhesion_energy}).

To account for capsomer-lipid excluded-volume interactions, a layer of 35 `excluder' beads, each with diameter $1.25\sigma$, is placed in the pentagon plane (Fig. \ref{fig:model}a). Excluders experience repulsive interactions with all lipid beads.
Since the mean location of the attractive interaction sites on adsorbed subunits is near the membrane midplane, the effective radius of the assembled capsid (not including the lipid coat) can be estimated from the distance between the attractors and the capsomer plane plus the capsid inradius (the radius of a sphere inscribed in a dodecahedron), which gives $R_\text{capsid}\approx15.3\sigma$.  As discussed below, this is smaller than any enveloped virus, and thus our results are qualitative in nature.

In this work we are motivated by viruses such as HIV, where expression of the capsid protein (GAG) alone is sufficient for the formation of budded particles \cite{Johnson2002}. Therefore we consider a model which does not include viral transmembrane proteins (spike proteins). We also do not consider how some viruses use cellular machinery to drive scission \cite{Baumgartel2011} as this process is virus-specific and depends on detailed properties of cellular proteins. For those viruses our model may elucidate the mechanisms leading up to the point of scission.

\subsection*{Simulations}
\protect\label{sec:simulations}
Simulations were performed on GPUs with a modified version HOOMD 0.10.1 \cite{Anderson2008,Nguyen2011a}. We modified the Andersen barostat \cite{Andersen1980} implementation to simulate the membrane at constant temperature and constant tension \cite{Reynwar2007} and to couple the barostat to rigid-body dynamics.
The membrane was coupled to the thermostat and barostat with characteristic times $\tau_T=0.4\tau_0$ and  $\tau_P=0.5\tau_0$ respectively, with $\tau_0$ the characteristic diffusion time for a lipid bead (defined below). The imposed tension was set to zero \cite{tensionNote}.

Each capsomer was simulated as a rigid body using the Brownian dynamics algorithm, which uses the (non-overdamped) Langevin equation to evolve positions and rigid body orientations in time \cite{Anderson2008, Nguyen2011a}.  To  approximate the rotational dynamics of globular proteins, we modified the rigid-body algorithm in HOOMD so that forces and torques arising from drag and random buffeting were applied separately and isotropically. Finally, the code was modified to update rigid-body positions according to changes in the box size generated by the barostat at each time step.

Matthews and Likos \cite{Matthews2013a} showed that hydrodynamic interactions (HI) between lipid particles can increase the rate of membrane deformation. However, given that the mechanisms of assembly and budding appeared to be similar in simulations which did not include HI, the timescales for protein diffusion and association are only qualitative in a coarse-grained model, and the large computational cost required to include HI in our more detailed model, we neglect HI in our simulations.

{\it Units.} We set the units of energy, length, and time in our simulations equal to the characteristic energy, size, and diffusion time for a lipid bead: $\epsilon_{0}$, $\sigma$ and $\tau_{0}$ respectively.
The remaining parameters can be assigned physical values by setting the system to room temperature, $T=300K$, and noting that the typical width of a lipid bilayer is around 5 nm, and the mass of a typical phospholipid is about 660 g/mol. The units of our system can then be assigned as follows: $\sigma=0.9$ nm, $m_{0}=220$ g/mol, $\epsilon_{0}=3.77 \times 10^{-21} \text{J}=227\text{g}\text{\AA{}}^{2}/\text{ps}^{2}\text{mol}$, and $\tau_{0}=\sigma\sqrt{m_{0}/\epsilon}=8.86$ ps. For each set of parameters, the results from four independent simulations were averaged to estimate the mean behavior of the system.

{\it Timescales.} The diffusion coefficient of capsomers in solution is $D\approx4.2 \sigma^2\tau_0$ while for capsomers adsorbed on the membrane $D\in [0.004,0.02]$ for $\ead$ ranging from 0.6 to 0.2. Thus timescales to diffuse by one capsomer diameter ($10\sigma$) are $\tau_\text{d}\approx25\tau_0$ for capsomers in solution and $\tau_\text{D} \in [500,2500]$ on the membrane.

{\it System.} To simulate an infinite membrane, periodic boundary conditions were employed for the lateral dimensions and a wall was placed at the bottom of the box. Thus, the capsomers remained below the membrane unless they budded through it.  To maintain a constant and equal ideal gas pressure above and below the membrane (despite the imbalance of capsomer concentrations), `phantom' particles were added to the system. These particles experienced excluded-volume interactions with the lipid head beads, and no other interactions.

For most simulations of inhomogeneous membranes the membrane contained $n=16,200$ lipids, including those belonging to the domain. An initial bilayer configuration was equilibrated and then placed with its normal along the $z$-axis in
a cubic box of side-length $L_x=L_y=90 \sigma$ and $L_z=100\sigma$.  For large domains ($r_\text{domain}>40$) the membrane contained $n=28,800$ lipids and the initial box size was $130\times130\times100 \sigma ^{3}$.
For most simulations of homogeneous membranes the bilayer contained $n=7,164$ lipids and the initial box size was $63.5\times63.5\times100 \sigma ^{3}$; additional simulations on larger membranes were performed to rule out finite size effects.

The capsomers were introduced in the box in two different ways, to understand how the rate of subunit translation and/or targeting to the membrane affects assembly.  The first set of simulations considered budding via quasi-equilibrium states, meaning that capsid proteins adsorb onto the membrane slowly in comparison to assembly and membrane deformation timescales. This scenario corresponds to the limit of low subunit concentration and a rate of subunit protein translation or targeting of subunits to the membrane which is slow in comparison to assembly. Specifically, each new capsomer was injected around 50$\sigma$ below the membrane midplane once all previously injected subunits were part of the same cluster. For other simulations, capsomers were injected one-by-one with an interval $\tinject$  until reaching a predefined maximum number of subunits. In the limit of $\tinject=0$, all capsomers were placed randomly at distances between 30 and 50$\sigma$ below the membrane at the beginning of the
simulation. For all simulations, the initial configuration had three free capsomers placed at 30 $\sigma$ below the membrane.

\section*{Results}

To simulate capsid protein and membrane dynamics on time- and length-scales relevant to assembly and budding, we use the models illustrated in Fig.~\ref{fig:model}a,b.
The physical mechanisms that control the formation and size of domains (with a typical size of 10-100 nm) in cell membranes  are poorly understood \cite{Lingwood2010,Kerviel2013,Parton2013}. To focus on the effect of a domain on assembly rather than its formation, we simulate a minimal heterogeneous membrane comprised of two lipid species, with interaction strengths that lead to phase separation
within
the membrane, with the minor species forming a circular domain (Fig.~\ref{fig:model}d). The bulk membrane and domain have the same bending coefficient and area per lipid (to focus on mechanisms other than curvature- or bending stiffness-sorting \cite{Baumgart2011}), but protein subunits preferentially partition into the domain. A complete listing of the interaction potentials is provided in Methods and SI Sec. ~\ref{sec:capsomer_model}.

%To understand the influence of membrane properties on assembly and budding,
We performed simulations for a range of subunit-membrane interaction strengths $\ead$, microdomain sizes $\rdomain$, microdomain line tensions $\ltension$, and timescales for subunit association to the membrane $\tinject$. All simulations were performed with a subunit-subunit interaction strength $\epsilon_{att}^v=4.0\epsilon_v$,
$\epsilon_{att}^e=2.0\epsilon_v$. While assembly can proceed in bulk under these conditions, in all simulations that we performed (for all values of $\tinject$) subunits adsorbed onto the membrane before assembling into any oligomer larger than a trimer. This behavior is consistent with enveloped viruses for which assembly in the cytosol is limited to small oligomers (e.g. HIV \cite{Ivanchenko2009}).  The results presented here correspond to long but finite simulation times, at which point assembly outcomes appeared roughly independent of increasing simulation time.  While these results need not necessarily correspond to equilibrium configurations, note that capsid assembly must proceed within finite timescales in {\it in vivo} or {\it in vitro} settings as well \cite{Hagan2013}.

\begin{figure}
\begin{center}
\includegraphics[viewport=0 0 590 361, width=0.48\textwidth]{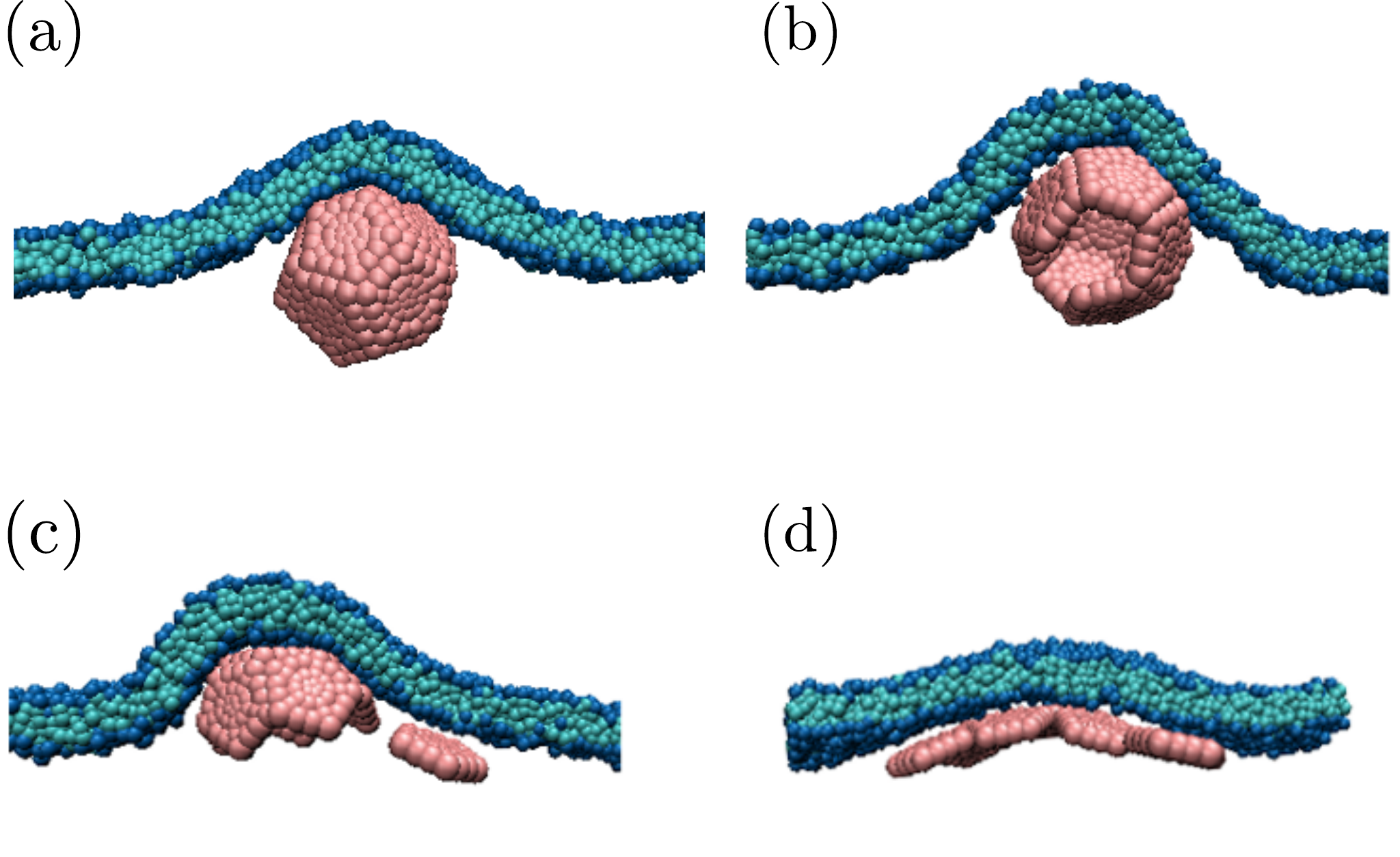}
% homogeneous.pdf: 590x361 pixel, 72dpi, 20.81x12.74 cm, bb=0 0 590 361
\end{center}
\caption{Typical end products for assembly on a homogeneous membrane as a function of subunit-membrane adhesion strength $\ead$. {\bf (a), (b)} Assembled but partially wrapped capsids for (a) $\ead=0.1\alpha$ and (b) $\ead=0.15\alpha$.  {\bf (c)} Assembly stalls at a half capsid for $\ead=0.2\alpha$. {\bf (d)} A deformed, open structure forms for $\ead=0.4\alpha$.
}
\label{fig:noraft}
\end{figure}

\subsection*{A homogeneous membrane introduces barriers to assembly}
Given that capsid proteins may be targeted to  the membrane rather than arriving by diffusion \cite{Balasubramaniam2011}, we have considered several modes of introducing subunits into our simulated system, as described in Methods.
We began by simulating assembly on a homogeneous membrane (a single species of lipid) (Fig.~\ref{fig:model}c) via quasi-equilibrium states, meaning that free subunits were injected into the system far from the membrane one-by-one, each after all previously injected subunits were assembled (see Methods). This scenario corresponds to the limit of low subunit concentration and a rate of subunit protein translation or targeting of subunits to the membrane which is slow in comparison to assembly.

 We found that assembly of membrane-absorbed subunits required large subunit-subunit interactions (as compared to those required for assembly in bulk solution), but that such subunits could undergo rapid nucleation on the membrane.  However, we found no sets of parameter values for which our model undergoes complete assembly and budding on a homogeneous membrane. In most simulations, assembly slows dramatically after formation of a half-capsid (six subunits). The nature of subsequent assembly depends on the adhesion strength. For  low adhesion strengths ($e_\text{ad}<0.2 \alpha$), assembly beyond a half-capsid occurs when particles detach from the membrane, sometimes leading to nearly completely assembled but  partially wrapped capsids (Fig.~\ref{fig:noraft} a,b). At intermediate adhesion strengths ($0.2\le e_\text{ad}/\alpha \le 0.4$)  particles do not readily dissociate from the membrane and assembly typically stalls at a half-capsids. Higher adhesion strengths ($\ead>0.4 e_\alpha$) yield deformed, open structures which cannot drive complete budding (Fig.~\ref{fig:noraft}d).

These results reveal that adsorption to a membrane has mixed effects on assembly.  Through dimensional reduction, membrane adsorption reduces the search space and thus can promote subunit-subunit collisions.  Furthermore, as shown in Matthews et al \cite{Matthews2012,Matthews2013a}, adsorption to the membrane can lead to high local subunit concentrations and thus reduce nucleation barriers. Similar effects occur during assembly on a polymer \cite{Kivenson2010,Elrad2010,Mahalik2012,Perlmutter2013,Hagan2013}.
However, assembly on the membrane also introduces new barriers to assembly.  First, formation of a completely enveloped capsid incurs a membrane bending free energy cost of $8\pi\kappa$, independent of capsid size \cite{Phillips2013}.  This free energy penalty must be compensated by subunit-subunit and subunit-membrane interactions.  In our model the subunit-membrane interactions do not promote membrane curvature, and thus large subunit-subunit interactions were required for assembly on the membrane. For these parameters nucleation also occurs in bulk solution if there is no membrane present (nucleation did not occur in bulk solution with a membrane present for any value of $\tinject$ because subunits adsorbed onto the membrane before undergoing nucleation). We also considered a model in which the surface of the subunit is curved (SI Fig.~\ref{fig:curved_capsomer}), so that subunit-membrane adsorption does promote local curvature. Interestingly, this model did not lead to improved assembly as compared to the flat subunits.

This surprising result and the
 frustrated assembly dynamics of half-capsid intermediates reveal additional kinetic and free energetic barriers to assembly, which are geometric in origin. For intermediates below half-size, the capsid-induced membrane curvature is positive everywhere, and further assembly requires only a small change in the angle of an approaching adsorbed subunit. On the other hand, assembly beyond half-size induces a neck characterized by negative curvature (SI Fig.~\ref{fig:geom_assembly}a); consequently, subunits approach the assembling partial capsid with orientations that are not conducive to association. Addition of such a subunit requires a large membrane deformation, which is energetically unfavorable for physically relevant values of the membrane bending rigidity and thus rare (SI Fig.~\ref{fig:geom_assembly}b).  Assembly therefore stalls or, in the case of weak adhesion energy, proceeds by detachment of subunits from the membrane leading to assembled
but partially wrapped
capsids. The stalled assembly states resemble the partially assembled states theoretically predicted by Zhang and Nguyen \cite{Zhang2008}, while the partially wrapped capsids are consistent with the metastable partially wrapped states found for a pre-assembled particle in our previous simulations \cite{Ruiz-Herrero2012}.
A second barrier arises because subunit-membrane interaction energies are reduced in regions where the membrane curvature is large on the length scale of the rigid subunit. This effect introduces a barrier to subunit diffusion across the neck (see the animation SI Video~\ref{fig:movie}), thus decreasing the flux of subunits to the assembling capsid.

As discussed below, the large magnitude of the membrane-induced barrier to assembly arises in part due to the small capsid size and relatively large subunits of our model. However, the barrier is intrinsic to assembly of spherical or convex polygonal structure on a deformable two-dimensional manifold and thus will exist for any such model.

\begin{figure}
\begin{center}
\includegraphics[viewport=0 0 1015 903, width=0.45\textwidth]{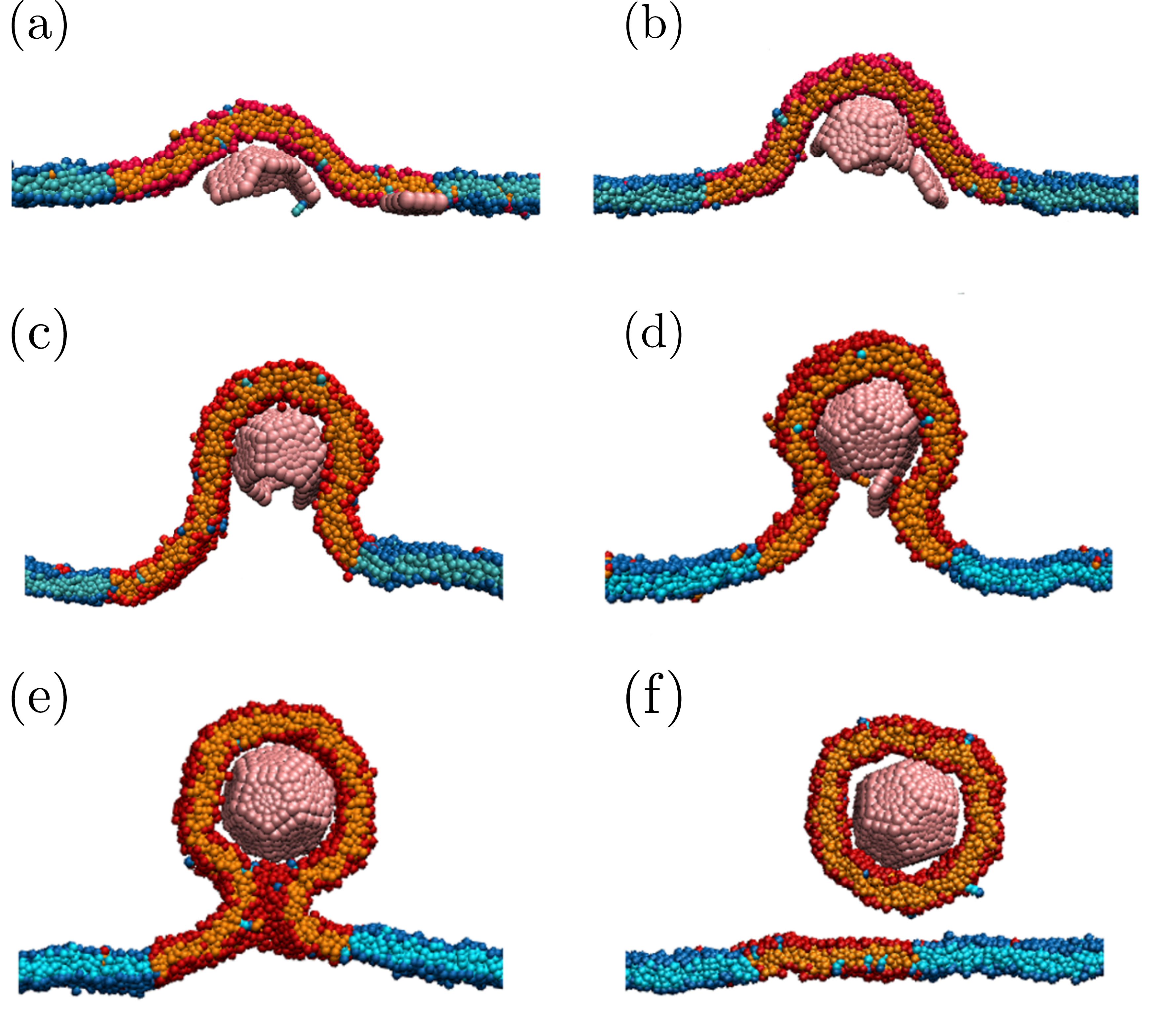}
% budding.pdf: 1015x903 pixel, 72dpi, 35.81x31.86 cm, bb=0 0 1015 903
\end{center}
\caption{Capsid assembly and budding from a domain.  2D slices of configurations at different times extracted from MD simulations for $\ead=0.4\alpha$, $r_\text{raft}=35\sigma$ and $\gamma=0.125\gamma_0$. The membrane wraps the growing capsid {\bf (a-d)} until the complete, enveloped capsid is connected to the rest of the membrane by a narrow neck {\bf (e)}. Finally, thermal fluctuations lead to fusion of the neck and the encapsulated capsid escapes from the membrane {\bf (f)}.
}
\label{fig:budding}
\end{figure}

\subsection*{Assembly and budding from a membrane microdomain}
\label{sec:capsomers_domain}
We next simulated assembly in the presence of a phase-separated membrane (Fig.~\ref{fig:model}d) to understand the effects of a membrane domain on assembly and budding. While there is some evidence that capsid proteins may induce the formation of lipid rafts \cite{Parton2013}, the mechanisms of lipid raft formation remain controversial. Here, we focus on the effect that the presence of a domain can exert on assembly and budding. We emphasize that we consider lipid-lipid interaction parameters for which the domain is flat and stable in the absence of capsid subunits (see SI Fig.~\ref{fig:line_tension}b); i.e., the domain line tension is insufficient to drive budding. We first consider budding in the quasi-equilibrium limit.

\begin{figure*}
\begin{center}
\includegraphics[width=0.98\textwidth]{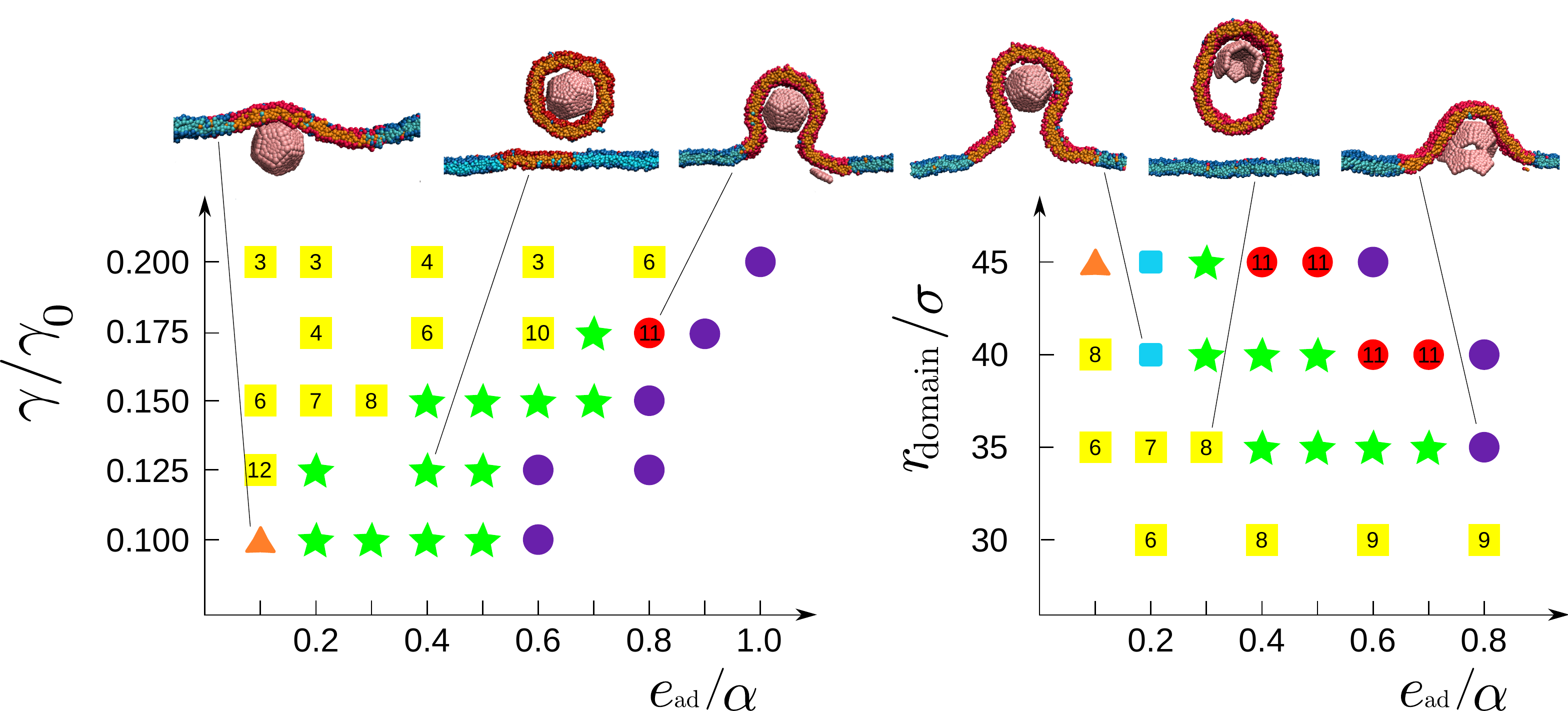}
\end{center}
\caption{Predominant end products from assembly simulations via quasi-equilibrium states with a membrane microdomain, as a function of adhesion strength $\ead$ and {\bf (left)}  line tension $\gamma$ with fixed domain radius $\rdomain=35\sigma$ and {\bf (right)} varying $\rdomain$ with fixed line tension $\gamma=0.15\gamma_0$, with $\gamma_0=13.4\kt/\sigma$ and $\alpha=-135.3\kt$. The most frequent outcomes are indicated as: complete assembly and budding (\textcolor{green}{$\bigstar$} symbols), budding of the entire domain before assembly completes, with the number indicating the typical partial capsid size upon budding %\footnote{The standard deviation in the partial capsid size we show inside the yellow squares is below 1}
(\textcolor{yellow}{$\blacksquare$} symbols), complete assembly but incomplete wrapping (\textcolor{orange}{$\blacktriangle$} symbols),  stalled assembly with wrapping (\textcolor{red}{$\CIRCLE$} symbols), complete assembly and wrapping without fusion of the neck (\textcolor{cyan}{$\blacksquare$} symbols), and malformed assembly (\textcolor{violet}{$\CIRCLE$} symbols).  Snapshots from simulations for the corresponding parameter sets are also shown. The complete distribution of outcome frequencies and assembly times are shown for some parameter sets in SI Fig.~\ref{fig:cumulative} and \ref{fig:assembly_time}.}
\label{fig:phase_diag1}
\end{figure*}

{\it Effect of line tension and adhesion energy.} Figure~\ref{fig:phase_diag1} (left) shows the predominant final system configurations as a function of $e_\text{ad}$ and $\gamma$  for fixed domain size $r_\text{domain}=35\sigma$, which is nearly twice the area required to wrap the capsid.
Moderate adhesion strengths and small line tensions lead to complete assembly and budding (Fig.~\ref{fig:budding}), meaning that:  12 subunits form a complete capsid, the capsid is completely wrapped by the membrane, and the membrane undergoes scission through spontaneous fusion of the neck to release the membrane-enveloped capsid. Because it requires a relatively large thermal fluctuation,  scission is characterized by long time scales. After scission, the portion of the domain not enveloping the capsid remains within the membrane.

Analysis of simulation trajectories identified several mechanisms by which the domain facilitates assembly. Firstly, partitioning of adsorbed proteins into the domain generates a high local subunit concentration. Secondly, the domain line tension promotes membrane curvature, since this reduces the length of the domain interface \cite{Lipowsky1993}. While this effect is not sufficient to drive domain curvature on an empty membrane for the parameters we consider, it facilitates membrane curvature around a partial capsid within the domain.  While the first two mechanisms could be anticipated based on existing theoretical knowledge, the simulations also identified a third mechanism that we did not anticipate. Namely,
 domains with sizes of order 2-4 times the area of a wrapped capsid promote long, shallow necks around assembly intermediates. While curvature energy favors capsid assembly in the domain interior, the line tension is minimized by a neck which extends to the domain interface. The relatively shallow curvature of such a neck greatly reduces the thermodynamic and kinetic barriers to assembly discussed in the previous section. Subunits diffuse readily across a long neck, and subsequent attachment to the assembling capsid incurs relatively small membrane deformation energies. The influence of the neck on subunit diffusion and association is illustrated by animations from assembly trajectories in SI Video \ref{fig:movie}.

Outside of optimal parameter values, we observe five classes of alternative end products. {\it (i)} For large values of the line tension $\gamma$, formation of a partial capsid triggers budding of the entire domain before assembly completes.  Under these parameters, the interfacial energy provides a driving force for budding of the entire domain, which is balanced by curvature energy in the absence of assembly \cite{Lipowsky1993}.  However, once the assembly of a partial capsid induces sufficient membrane curvature, the interfacial energy dominates and the domain buds. Within this region, the number of subunits found within the budded domain and the threshold value of the surface tension required for entire-domain budding increase with  $e_\text{ad}$.  This trend arises because stronger subunit-membrane adhesion leads to tight wrapping of intermediates and thus larger assemblages are required for the induced curvature to propagate to the domain interface.

\begin{figure}
\begin{center}
\includegraphics[bb=0 0 782 750, width=0.5\textwidth]{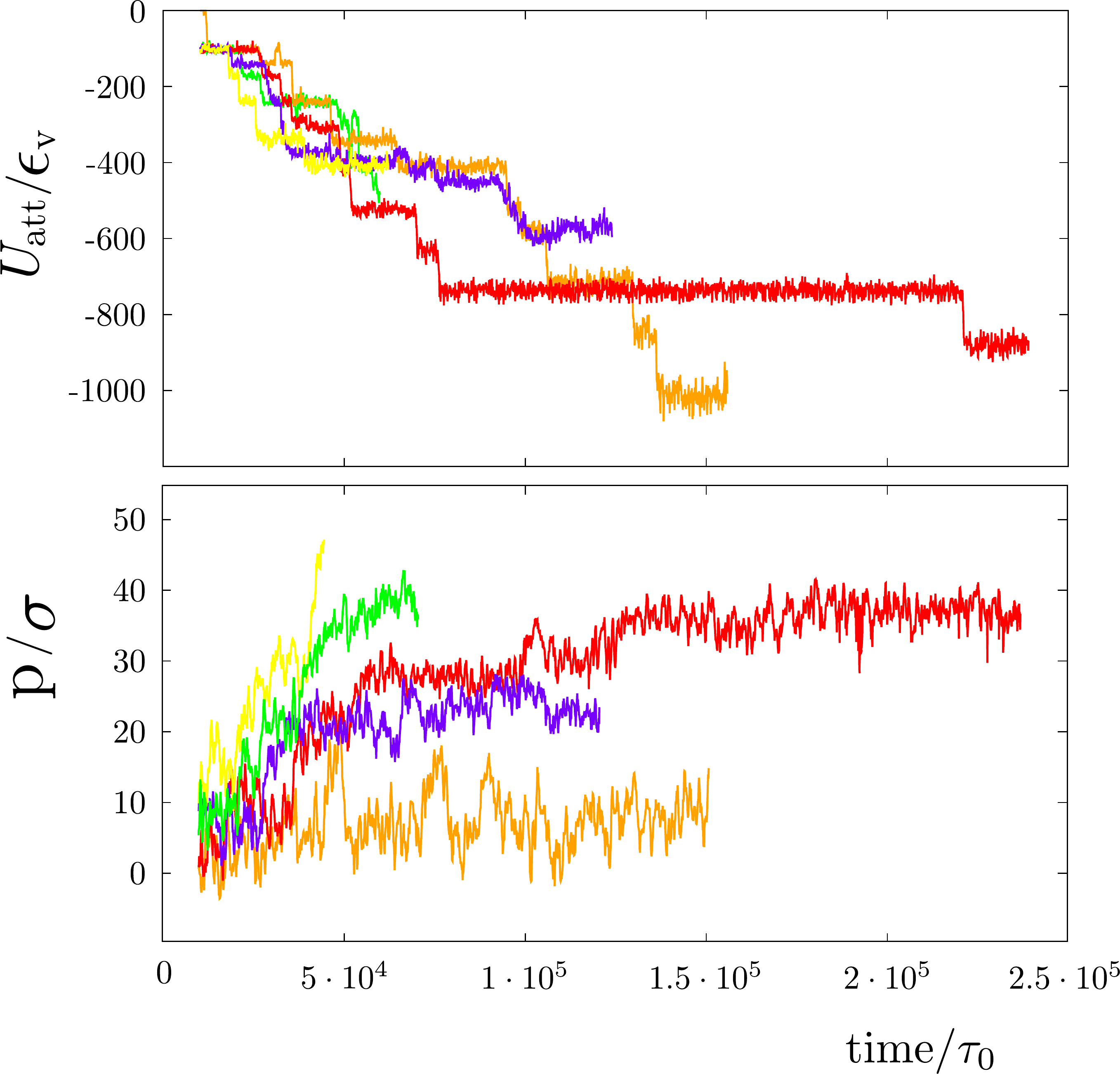}
% traj.pdf: 782x750 pixel, 72dpi, 27.59x26.46 cm, bb=0 0 782 750
\end{center}
\caption{Total subunit-subunit attractive interaction energy (top) and amplitude of membrane deformation, measured by the capsid penetration, $p$, (bottom) as a function of time for a trajectory leading to each type of outcome described in the main text. The capsid  penetration $p$ is measured as the distance between the top of the capsid and the center of mass of the membrane. The color code represents the outcome type and follows the same format as in Fig.~\ref{fig:phase_diag1}: successful assembly (green), budding of a partial capsid (yellow), complete assembly but incomplete wrapping (orange), stalled assembly with wrapping (red) and malformed assembly (violet).}
\label{fig:trajectories}
\end{figure}

{\it (ii)} For small $\gamma$  and $\epsilon_\text{ad}$, the capsid assembles but wrapping is incomplete. Here the subunit-membrane adhesion energy is insufficient to compensate for the membrane bending energy cost associated with wrapping.  {\it (iii)}
For larger-than-optimal adhesion strengths, the membrane wraps the assembling capsid tightly with a short neck. As discussed in the previous section, the high negative curvature associated with a short neck inhibits association of the final subunit leading to stalled, incomplete assembly. {\it (iv)} For large $e_\text{ad}$, subunit-membrane adhesion energy dominates over subunit-subunit interactions leading to mis-assembled structures. Finally, {\it (v)} at other domain sizes (Fig.~\ref{fig:phase_diag1} right) we observe configurations in which the capsid is completely wrapped, but the neck does not undergo fusion.
To illustrate the timescales, interactions, and coupling between assembly and membrane configurations, the total subunit-subunit attractive interaction energy and the magnitude of membrane deformation are plotted as a function of time for a trajectory leading to each type of outcome in Fig.~\ref{fig:trajectories}.

%\commentbis{I added a figure in the SI, the last one, showing a typical process of flat aggregates formation. Consider refering to it or removing it if you think it is not important.}
%I suggest we remove it

{\it Effect of domain size.}
The dependence of assembly and budding on the domain radius $\rdomain$  for constant line tension $\gamma=0.15\gamma_0$ is shown in Figure~\ref{fig:phase_diag1} (right).  There is an optimal domain size with about twice the area of a wrapped capsid ($35 \sigma \lesssim \rdomain \lesssim 40 \sigma$) that leads to robust assembly and budding over a broad range of adhesion energies $\ead$. For smaller domains, low values of adhesion lead to budding of the entire domain before assembly completes. This result was unexpected -- in the absence of protein assembly, line tension triggers budding {\it above} a threshold domain size; smaller domains are stable because bending energy dominates over interfacial energy \cite{Lipowsky1993}.  However, we find here that partial capsid intermediates stabilize membrane deformation over an area proportional to their size, and thus drive budding within domains {\it below} a threshold size. On the other hand, for larger than optimal domains the assembling capsid only deforms a
fraction of the domain, and the domain interface does not promote a long neck or curvature around the capsid.  The behavior of such a domain is therefore comparable to that in a homogeneous membrane.

%We anticipate that there is a threshold domain size above which complete assembly and wrapping %would not occur for this model, but we were not able to confirm this possibility because simulations %with $\rdomain>50\sigma$ were computationally prohibitive. \commentbis{should we introduce the %assembly times we show in the last figure of SI?}

\begin{figure}
\begin{center}
\includegraphics[viewport=0 0 380 299, width=0.45\textwidth]{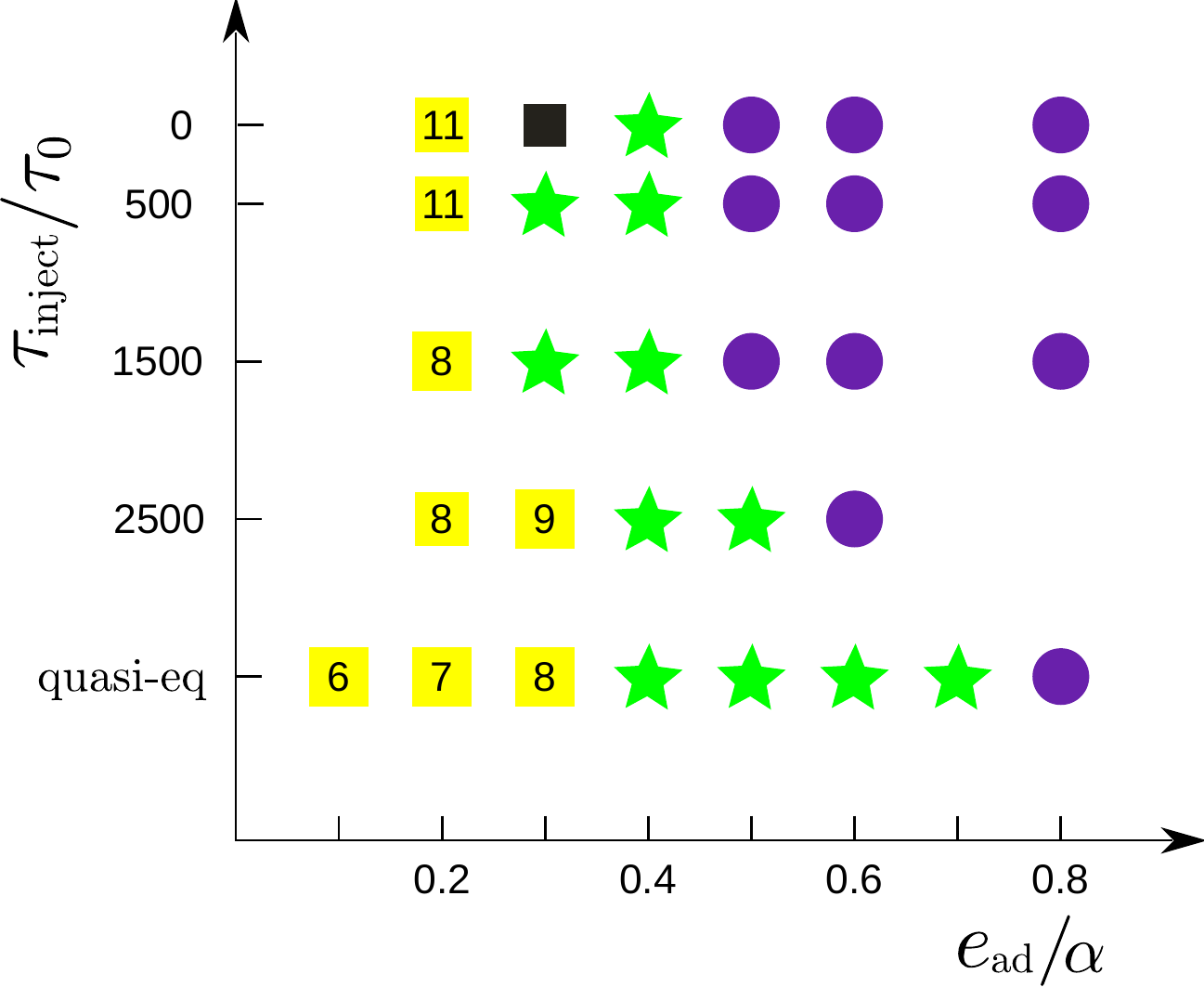}
% phasediagdynamics_soft.pdf: 380x299 pixel, 72dpi, 13.41x10.55 cm, bb=0 0 380 299
\end{center}
\caption{Predominant end products as a function of the subunit injection timescale $\tinject$ and the adhesion strength $\ead$ are shown for a domain with $r_\text{domain}=35\sigma$ and $\gamma=0.15\gamma_0$.  The most frequent outcome is shown for every set of parameters. Symbols are defined as in Figure~\ref{fig:phase_diag1} except for \textcolor{black}{$\blacksquare$} symbols, which denote budding of the whole domain with a malformed capsid inside. Alternative outcomes observed at some parameter sets are documented in SI Fig.~\ref{fig:phase_diagdyn_asterisks}.}
\label{fig:phase_diagdyn}
\end{figure}

{\it Effect of subunit adsorption timescale.}
In the quasi-equilibrium simulations discussed so far, the assembly outcomes were determined by the relative timescales of membrane deformation and partial capsid annealing.   To determine the effect of the subunit adsorption timescale, we characterized the system behavior for subunit injection timescales $\tinject$ (see Methods) between the quasi-equilibrium limit and 0, where all subunits were introduced at the inception of the simulation (Fig. \ref{fig:phase_diagdyn}). We set $r_\text{domain}=35\sigma$, which led to relatively robust budding in the quasi-equilibrium limit.

The predominant end products are shown as a function of the adhesion strength and the subunit injection timescale in Figure \ref{fig:phase_diagdyn}. We see that the qualitative behavior is independent of the injection timescale; for all injection rates there is range of intermediate adhesion strengths around $e_\text{ad}=0.4\alpha$ for which complete assembly and budding is observed. However, as the injection timescale decreases, both the lower and upper bounds of this optimal range shift to weaker adhesion energies.  At weak adhesion energies the increased frequency of subunit binding promotes complete assembly and budding by reducing the overall assembly timescale below the timescale for budding of the entire domain.

Stronger-than-optimal adhesion energies tend to result in malformed assemblages (SI Fig.~\ref{fig:kinetic_traps}b) at the lower injection timescales.
This result can be understood from previous studies of assembly into empty capsids or around polymers \cite{Hagan2006,Elrad2010,Hagan2011,Hagan2013,Grant2011,Nguyen2007,Rapaport2010} -- higher adhesion energies lead to an exponential increase in the timescale for annealing of partial capsid configurations; kinetic traps occur when annealing timescales exceed the subunit binding timescale.
The ultimate fate of these large aggregates depends on the adhesion energy $\ead$. For smaller-than-optimal adhesion energies, assemblages are loosely wrapped and the entire domain undergoes budding once the assemblage reaches a threshold size (e.g. SI Fig. \ref{fig:kinetic_traps}b). For larger $\ead$, malformed aggregates are tightly wrapped by the membrane and remain attached by a neck (e.g. SI Fig. \ref{fig:kinetic_traps}a). The shortest injection timescales and largest adhesion energies we investigated lead to large flat aggregates that do not bend the membrane (Fig. \ref{fig:kinetic_traps}c), or partial capsids emerging from a flat aggregate  (SI Fig. \ref{fig:kinetic_traps}d). Finally, we note that as the subunit injection timescale is decreased, the diversity of outcomes at a given parameter set increases and thus the yield of budded well-formed capsids decreases (SI Fig. ~\ref{fig:phase_diagdyn_asterisks}).

{\it Effect of subunit copy number.} We found that the dynamics is qualitatively similar when excess subunits are included in the simulation. For example, we performed simulations on systems with with 19 capsomers,  about 60\% more than needed for capsid formation.  For an injection timescale of $\tinject=500\tau_0$, the behavior is similar to the small $\tinject$ results discussed above, except that subunits on the periphery of an  assembling capsid typically form flat aggregates that can hinder budding (SI Fig. \ref{fig:budplenty}). For adhesion strengths between 0.3 to 0.4 $\epsilon_0$ budding is observed (SI Fig. \ref{fig:budplenty}), whereas larger values of $\ead$ lead to the forms of kinetic traps discussed above.

\section*{Discussion}

Our simulations demonstrate that, while a fluctuating membrane can promote assembly through dimensional reduction, it also introduces barriers to assembly by limiting the diffusion and orientational fluctuations of adsorbed subunits. These barriers, which are not present for assembly in bulk solution \cite{Hagan2013}, can engender metastable partially assembled or partially budded structures. While  barrier heights may depend on the specific membrane and protein properties (see below), their existence is generic to the assembly of a curved structure on a deformable surface. We find that assembly from a membrane microdomain can substantially reduce the effect of these barriers, which could partly account for the prevalence of enveloped viruses that preferentially bud from lipid rafts or other membrane microdomains.

As a first exploration of the relationship between membrane domain structure and budding, we considered a minimal model for a microdomain, which accounts only for preferential partitioning or targeting of capsid proteins within the domain. Our simulations identified two effects by which such a domain can promote assembly and budding:  (i) Generating a high local concentration of adsorbed subunits and (ii) Decreasing membrane-associated barriers to assembly by lengthening the neck around the budding capsid. While the first effect could be anticipated from standard reaction diffusion analysis, the second arises from the complex interplay between domain line tension and the geometry of a bud.  Importantly, the predicted effects are sensitive to the domain size (Fig.~\ref{fig:phase_diag1}), with an optimal domain size on the order of 2-3 times the area of a wrapped capsid. Smaller domains lead to budding before completion of assembly, whereas facilitation of budding becomes ineffective when the domain radius
becomes large in comparison to the capsid size. Importantly, these predictions can be directly tested by  {\it in vitro} experiments in which capsid proteins assemble and bud from artificial phospholipid vesicles with domains of varying sizes.

Finally, we consider the limitations of the model studied here. Although extending the model to relax these limitations is beyond the scope of the present work, doing so in future investigations could further elucidate the factors that control assembly and budding. The effective diameter of our enveloped $T{=}1$ capsid is about 28 nm, while the smallest enveloped viruses found in nature have diameters of 40-50 nm (e.g. \cite{Jones2003}). Although the relationship between particle size and budding has been explored in detail for preassembled nucleocapsids or nanoparticles (e.g. \cite{Ruiz-Herrero2012,Yue2011, Ginzburg2007}), our simulations here have identified new factors that control simultaneous assembly and budding. During assembly of a larger capsid, each subunit would individually comprise a smaller fraction of the total capsid area and thus would incur a smaller increment of membrane deformation energy when associating with the capsid. Similarly, intra-subunit degrees of freedom could allow subunit distortions that would facilitate diffusion across the neck.  However, note that such distortions would be free energetically unfavorable and thus would still impose a free energy barrier. We also note that the potential used for the subunit-membrane interaction in this work does not represent local distortions of the lipid hydrophobic tails resulting from insertion of a hydrophobic group. These distortions would most likely promote local negative curvature and thus might inhibit formation of membrane curvature; however, they could induce membrane-mediated attractions between subunits.  Given the qualitative nature of subunit-subunit interactions in our model, we do not expect these effects to qualitatively change the results.

While our model demonstrates three mechanisms by which a domain can promote membrane deformation, the effect of lipid and protein compositions within microdomains on membrane bending rigidity and spontaneous curvature could have additional effects \cite{Baumgart2011}. Similarly, for some viruses important roles are played by recruitment of additional viral proteins \cite{Solon2005}, other cellular factors that create or support membrane curvature \cite{McMahon2005, Doherty2009, Hurley2010}, and cytoskeletal machinery that actively drives budding (e.g. \cite{Taylor2011, Gladnikoff2009, Welsch2007, Balasubramaniam2011}). While these results can be systematically incorporated into the model, our current simulations provide an essential starting point to understand how microdomains facilitate budding and, through comparison with experiments, to identify the critical steps in budding.

\section*{Acknowledgments}
This work was supported by Award Number R01GM108021 from the National Institute Of General Medical Sciences, NSF-MRSEC-0820492, MODELICO Grant (S2009/ESP-1691) from Comunidad Aut\'onoma de Madrid, and FIS2010-22047-C05-01 Grant from Ministerio de Ciencia e Innovaci\'on de Espa\~na. Computational resources were provided by the National Science Foundation through XSEDE computing resources (Longhorn and Keeneland) and the Brandeis HPCC.

\section*{SUPPORTING MATERIAL}
Supporting material is available for this article.

\section*{SUPPORTING CITATIONS}
References \cite{Cooke2005,Weeks1971,Grest1986} appear in the Supporting Material.
% Here references are directly included this tex file.
% But we can generate reference list from bibliography database
% Compile and format the bibliography (bj_bibtex_template.bib BibTeX
% file must be present in the document directory)

%The source file for this document is called
%\emph{biophys\_latex\_template.tex}.  Apart from this \LaTeX\ file, you
%will also need the bibliography file, the \BibTeX\ style file, and the
%EPS and PDF figure files.

%See the bibliography file \emph{bj\_bibtex\_template.bib} for the
%literature data.  It was mostly generated from the saved
%text-formatted PubMed entries using the \emph{med2bib} program and
%edited by the \emph{tkbibtex} or directly in the \emph{emacs} editor.

%The \emph{biophysj.bst} file is a \BibTeX\ style file that contains
%information about the format required by Biophysical Journal for the
%list of references.

% \bibliography{library2}

 %Bibliography style (requires the style file biophysj.bst in the
% document directory)
\bibliographystyle{biophysj}

% Figure legends
%%Automatically it will add the figure legends  and table legends as a list by below command

\newpage

\newpage
\onecolumn 
\appendix{} 
\renewcommand\thefigure{S\arabic{figure}} 
\renewcommand{\theequation}{S\arabic{equation}}
\renewcommand{\thesection}{S\arabic{section}}

\chapter{SUPPORTING INFORMATION}
\setcounter{figure}{0} 
\section{Methods}

\subsection{The membrane model}
\label{sec:raft_model}

We model the amphiphilic lipids comprising the membrane with a coarse grained implicit solvent model from Cooke et al \cite{Cooke2005}, in which each amphiphile is represented by one head bead and
two tail beads that interact via WCA potentials \cite{Weeks1971}, equation (\ref{eq:wca})

\begin{equation}
V_\text{rep}(r)=\left\{\begin{array}{cr}
4 \epsilon_{0} \left[ \left( \frac{b}{r}\right) ^{12}-\left( \frac{b}{r}\right) ^{6}+\frac{1}{4}\right] &;r\leq r_\text{c} \\
0 & ;r>r_\text{c}
\end{array} \right.
\label{eq:wca}
\end{equation}
with $r_\text{c}=2^{1/6}b$ and $b$ is chosen to ensure an effective cylindrical lipid shape: $b_\text{head-head}=b_\text{head-tail}=0.95\sigma_0$ and $b_\text{head-tail}=\sigma_0$, where $\sigma_0$ will turn out to be the typical distance between beads within a model lipid molecule.

The beads belonging to a given lipid are connected through FENE bonds ( Eq.~(\ref{eq:bond})) \cite{Grest1986} and the linearity of the molecule is achieved via a harmonic spring with rest
length $4\sigma$ between the first and the third bead, Eq.~(\ref{eq:bend})
\begin{equation}
 V_\text{bond}(r)=-\frac{1}{2}\kappa_\text{bond} r_\text{cut}^{2} \ln \left[ 1-\left( r/r_\text{cut}\right) ^{2}\right]
\label{eq:bond}
\end{equation}

 \begin{equation}
V_\text{bend}(r)=\frac{1}{2}\kappa_\text{bend}\left( r-4\sigma_{0} \right)^{2}
\label{eq:bend}
 \end{equation}

\begin{figure}
\begin{center}
\includegraphics[viewport=0 0 852 340, width=0.8\textwidth]{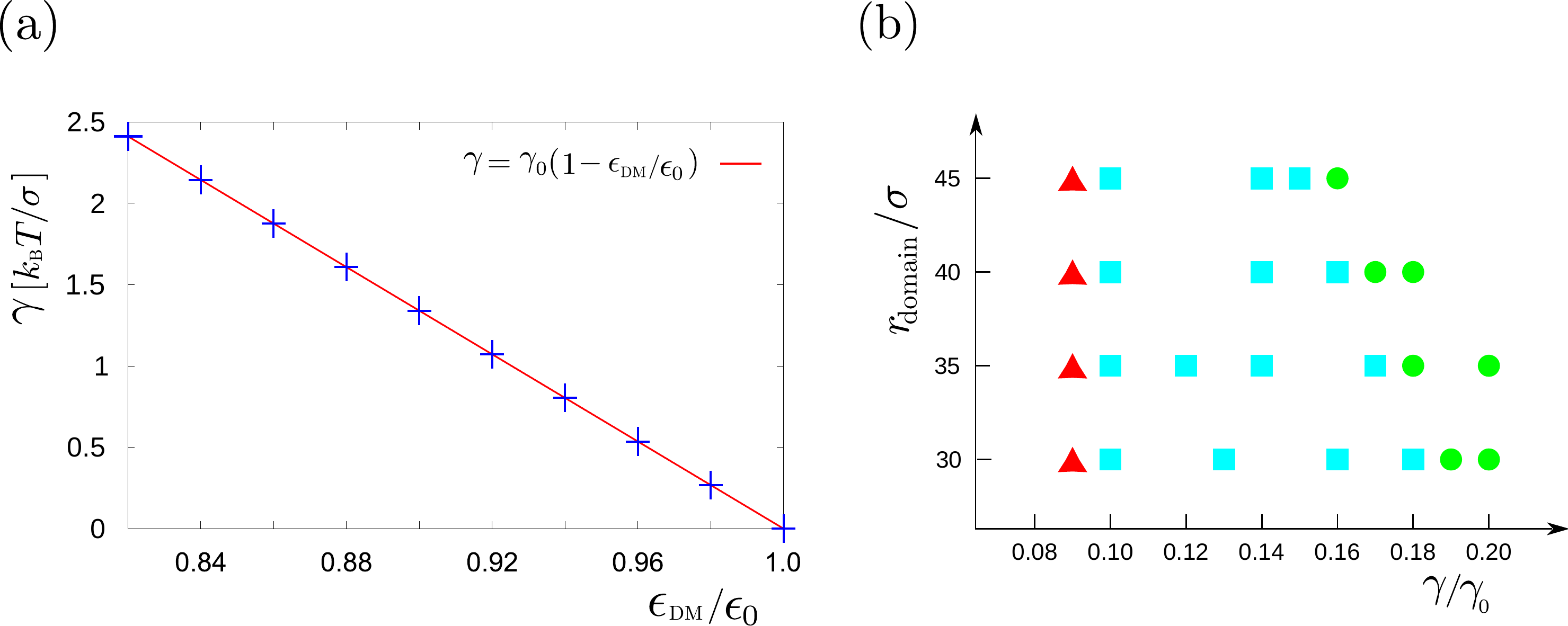}
% domain_dyn.pdf: 852x340 pixel, 72dpi, 30.06x11.99 cm, bb=0 0 852 340

\end{center}
\caption{{\bf (a)}The line tension (\textcolor{blue}{$+$} symbols) as a function of the interaction strength between particles belonging to different domains $\epsilon_\text{DM}$, calculated from the difference between the stress components normal and tangent to the interface \cite{Smith2007}. The solid line is a linear fit the data (Eq.~(\ref{eq:line_tension})). {\bf (b)}  Phase diagram of the domain behavior as a function of the domain radius and the line tension $\gamma$ obtained from Molecular Dynamic simulations of a membrane with a domain at $\kt=1.1\epsilon_0$. The possible outcomes are indicated as: domain dissolution (\textcolor{red}{$\blacktriangle$}), domain in equilibrium with the membrane (\textcolor{cyan}{$\blacksquare$}), and spontaneous budding of the whole domain (\textcolor{green}{$\CIRCLE$}). }
\label{fig:line_tension}
\end{figure}

Since this is an implicit solvent model, hydrophobicity is represented by an attractive interaction, equation (\ref{eq:hydrophobicity}), between all tail beads.  The molecules belonging to the domain are labeled \textit{D}, while those forming the rest of the membrane are referred as \textit{M}. The interaction between molecules of the same type is the same, but the strength of the effective hydrophobic interaction between molecules of different type is lower:
\begin{equation}
 U_\text{hydro}^{ij} (r)=\left\{ \begin{array}{ccc}
 -\epsilon_{ij} &;& r<r_\text{c} \\
-\epsilon_{ij} \cos^{2} \frac{\pi (r-r_\text{c})}{2\omega_\text{c}} &;&r_\text{c}\leq r \leq r_\text{c}+\omega_\text{c} \\
0 & ; & r>r_\text{c}+\omega_\text{c}
\end{array} \right.
\label{eq:hydrophobicity}
\end{equation}
where the interaction between the molecules of the same type is given by $\epsilon_\text{DD}=\epsilon_\text{MM}=\epsilon_{0}$,  and the cross term, $\epsilon_\text{DM}$, is a parameter that controls the strength of the line tension between domains. Varying $\epsilon_\text{DM}$ from 0 to $\epsilon_0$ tunes the line tension, from a large value to 0. The energy of the domain border is proportional to the line tension and the domain perimeter equation (~\ref{eq:domain_energy})
\begin{equation}
e_\text{domain}=2\pi r_\text{domain}\gamma
\label{eq:domain_energy}
\end{equation}
where $r_\text{domain}$ is the domain radius and $\gamma$ the line tension. We estimated the energy stored in the edge  from the difference between the stress components normal and tangent to the interface \cite{Smith2007}, and obtained a linear relation between the line tension and the attractive interaction strength between lipid molecules of different type (Fig.~\ref{fig:line_tension}a, Eq.~(\ref{eq:line_tension})):
\begin{equation}
\gamma=\gamma_0(1-\epsilon_{DM}/\epsilon_0)
\label{eq:line_tension}
\end{equation}
with $\gamma_0=13.4\kt/\sigma$ for $\omega_c=1.5\sigma$. The membrane phase behavior as a function of the line tension is shown in (Fig.~\ref{fig:line_tension}b). For small values of $\gamma$ the membrane composition is homogeneous, for intermediate values a phase-separated membrane is stable, while for high values of $\gamma$ interfacial tension drives budding of the entire domain.

This model allows the formation of bilayers with physical properties such as fluidity, area per molecule and bending rigidity that are easily tuned via
$\omega_{c}$. Moreover, diffusivity within the membrane, density, and bending rigidity are in good agreement with values of these parameters measured for  biological membranes\cite{Cooke2005} .

\subsection{The virus model}
\label{sec:capsomer_model}

Our model for capsid assembly is based on the model for $T{=}1$ capsids developed by Wales \cite{Wales2005}, but has been extended to allow for physically realistic interactions with the membrane. Our capsid subunit is a rigid body with a pentagonal base and radius of $r_\text{pentamer}=5 \sigma$ formed by 15 attractive and 10 repulsive interaction sites.  While the original model \cite{Wales2005} contained 5 attractive and 5 repulsive sites, the new sites that we have added (Figs.~\ref{fig:fullcapsomer} and ~\ref{fig:capsomer_model}) are necessary to describe assembly on a fluctuating surface. The effects of their inclusion are shown in the following sections.

\subsubsection{Attractor sites}

Subunit assembly is mediated through a Morse potential between `attractor' pseudoatoms located in the pentagon plane, with one located at each subunit vertex and 2 along each edge. Attractions occur between like attractors only, meaning that there are vertex-vertex and edge-edge attractions, but no vertex-edge attractor interactions, equation (\ref{eq:attractors})
\begin{eqnarray}
U_{att}&=&\epsilon_\text{att}^\text{v}\sum_{m}\sum_{n}\left( e^{\rho(1-R_{mn}/R_\text{e})}-2\right)e^{\rho(1-R_{jk}/R_\text{e})} \nonumber \\
&+& \epsilon_\text{att}^\text{e}\sum_{p}\sum_{q}\left( e^{\rho(1-R_{pq}/R_\text{e})}-2\right)e^{\rho(1-R_{lm}/R_\text{e})}
\label{eq:attractors}
\end{eqnarray}
where $\epsilon_\text{att}^\text{v}$ is the interaction strength between vertex sites, $\epsilon_\text{att}^\text{e}$ is the interaction strength between edge sites, $R_{mn}$ is the distance between sites \textit{m} and \textit{n},with  \textit{m} running over the attractor sites on the vertices of the first capsomer , and \textit{n} running over the vertices on the second.  $R_{pq}$ is the analogous distance between the  edge sites on each of the capsomers. $R_\text{e}$ is the equilibrium pair distance and $\rho$ defines the range of the interaction.

\begin{figure}
\begin{center}
\includegraphics[viewport=0 0 981 510,width=0.45\textwidth]{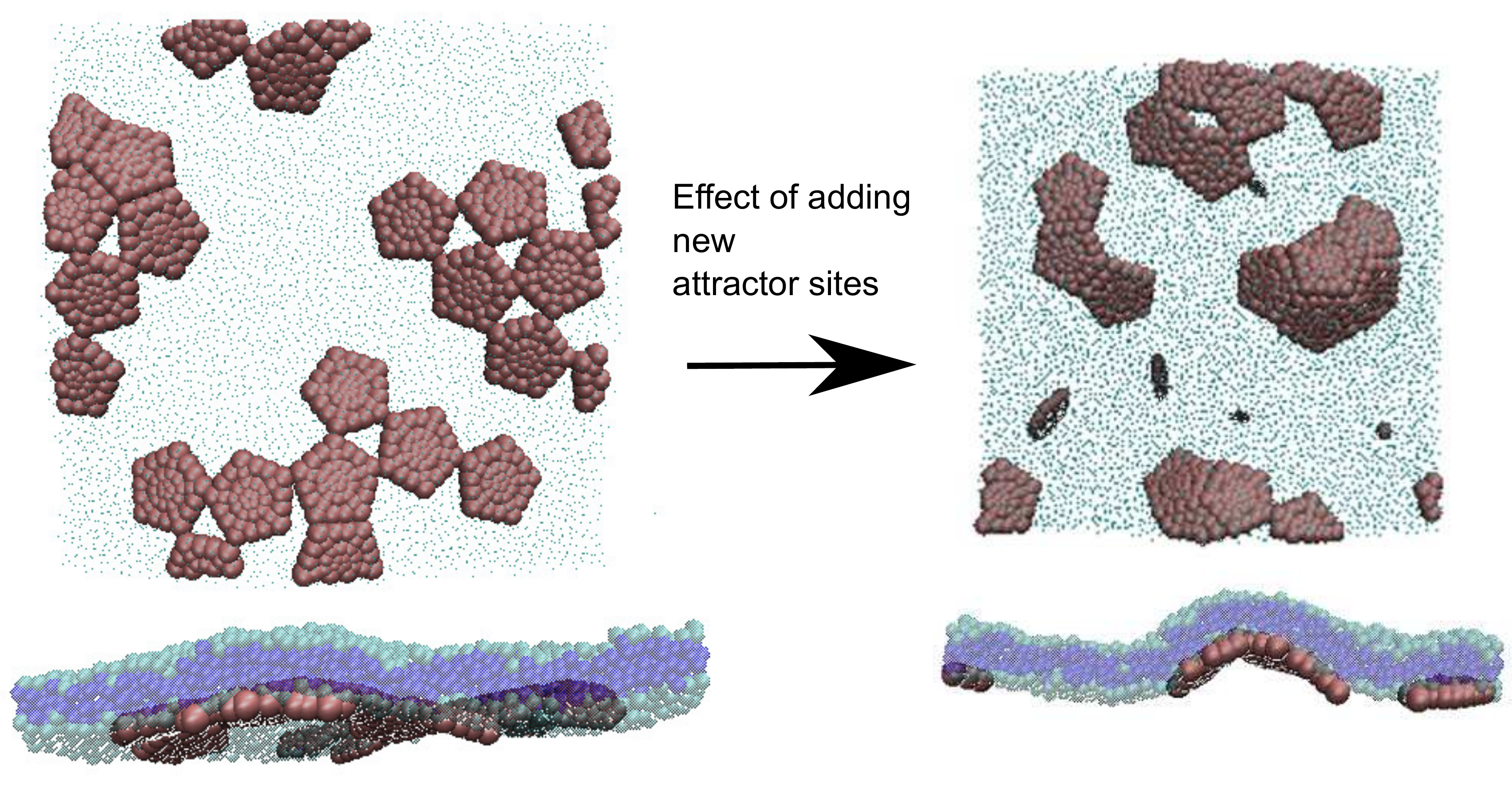}
% attractoseffect.pdf: 981x510 pixel, 72dpi, 34.61x17.99 cm, viewport=0 0 981 510
\end{center}
\caption{Effect of adding new attractors. While the original model  capsomers tended to form defective, flat assemblages on the membrane {\bf (left)}, the extended model capsomers assemble with the correct curvature and subunit-subunit interaction geometries {\bf (right)}.}
\label{fig:newattractors_effect}
\end{figure}

In comparison to the original model \cite{Wales2005} the additional attractive sites provide a stronger driving force for formation of structures with the lowest energy face-face angles and thus provide additional thermodynamic stabilization of the lowest energy dodecahedron capsid structure.  This increased stabilization of the curved, icosahedral shape is necessary to compete with membrane bending energy which favors flat aggregates. Although we did observe assembly on the membrane with the original model for carefully tuned parameters, the improved model is much more robust, meaning that it undergoes assembly over a much wider range of parameter values (Fig. \ref{fig:newattractors_effect}).

\subsubsection{Repulsive sites}

\begin{figure}
\begin{center}
\includegraphics[viewport=0 0 1030 919,width=0.4\textwidth]{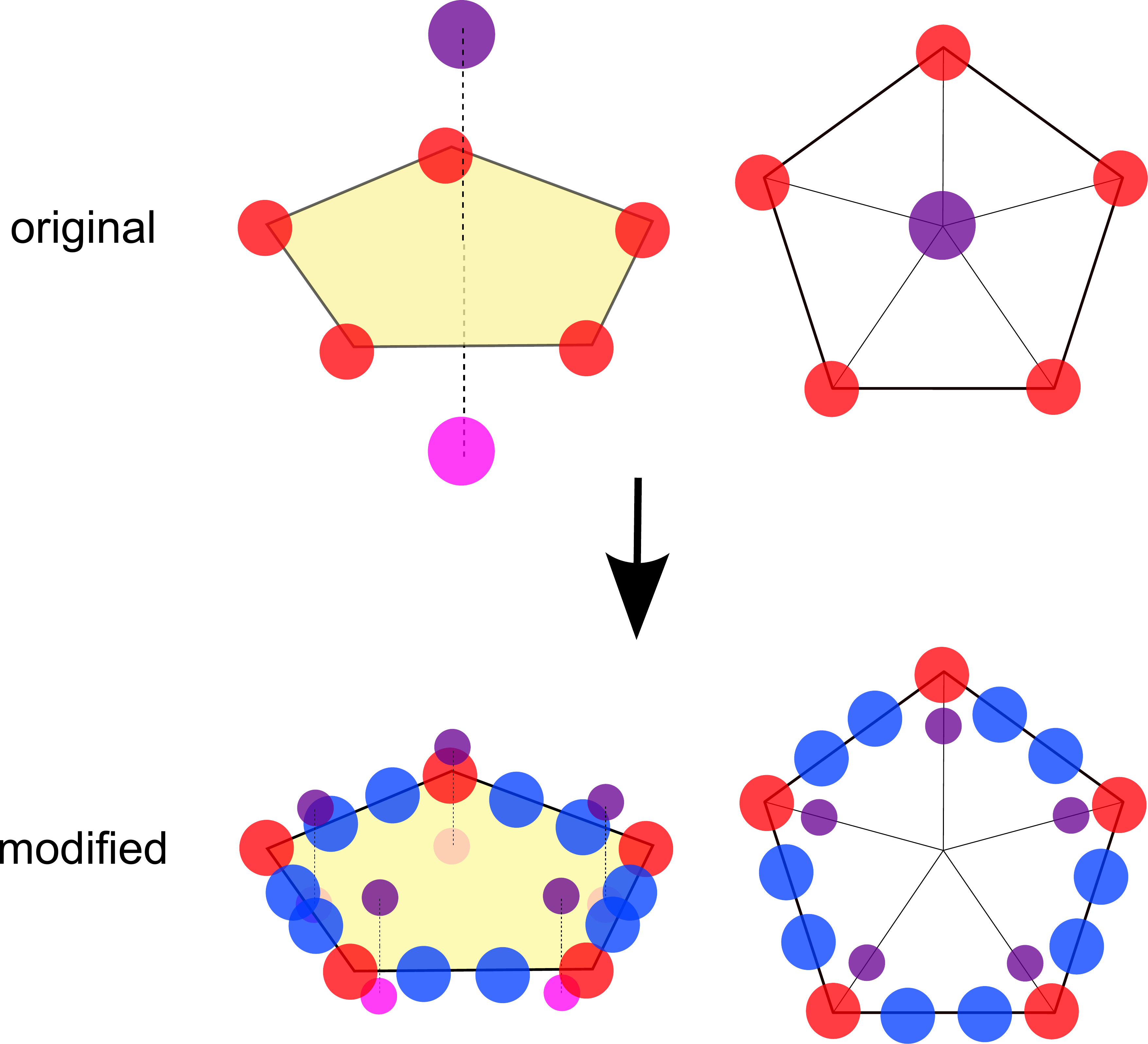}
%\includegraphics[width=0.4\textwidth]{./capsomerdiff.pdf}
% capsomerdiff.pdf: 1030x919 pixel, 72dpi, 36.34x32.42 cm, viewport=0 0 1030 919
\end{center}
\caption{Comparison of the original model \cite{Wales2005} and our extended model. Blue circles represent the new attractive sites on the capsomer edges, and the violet and magenta circles denote the new repulsive sites.}
\label{fig:capsomer_model}
\end{figure}

\begin{figure}
\begin{center}
\includegraphics[viewport=0 0 769 254, width=0.6\textwidth]{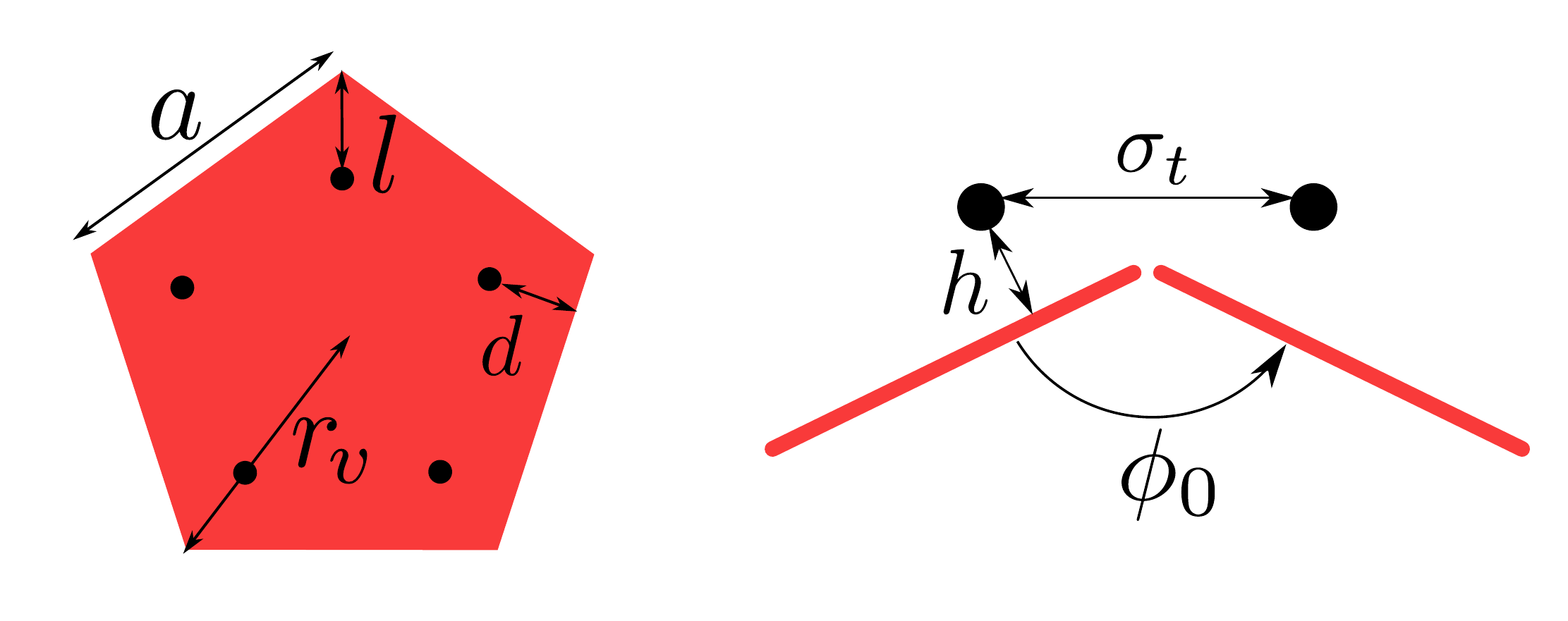}
% definition_sigma.pdf: 769x254 pixel, 72dpi, 27.13x8.96 cm, bb=0 0 769 254
\end{center}
\caption{Capsomer geometry in the extended subunit model. {\bf(Left):} a top view of a capsomer of radius $r_\text{v}$ and edge length $a$. The projections of the new repulsive sites on the capsomer plane lie on each of the pentamer radii, at distances $l$ from the nearest vertex and $d$ from the pentamer edge. Their distances from the capsomer plane, $h$, and $l$  keep the same proportions as those in the original model. {\bf(Right):} Geometry of capsomer-capsomer binding. For two adjacent pentamers in a complete capsid, the distance between two opposite repulsive sites is $\sigma_\text{t}$ and the equilibrium angle is $\phi_0$}
\label{fig:geometry_sigma}
\end{figure}

The 10 repulsive interaction sites are separated into 5 `top' and 5 `bottom' sites, which are arranged symmetrically above and below the pentagon plane respectively, so as to favor a subunit-subunit angle consistent with a dodecahedron (116 degrees). They are at distance $h$ from the capsomer plane, and their projections on that plane lie on each of the pentamer radii, at a distance $l$ to the corner. The ratio $h/l$ is the same as in the original model (Fig. \ref{fig:capsomer_model} and Fig. \ref{fig:geometry_sigma}). The interaction potential between top and bottom sites on two capsomers is similar to that in the original model but extended to all the sites:
\begin{equation}
U_{rep}=\epsilon _{rep}\sum_{i=1}^{5} \sum_{j=1}^{5}\left( \frac{\sigma_\text{t}}{R_{ij}}\right)^{12}+\epsilon_\text{rep} \sum_{m=1}^{5} \sum_{n=1}^{10}\left( \frac{\sigma_\text{b}}{R_{mn}}\right)^{12}
\label{eq:repulsors}
\end{equation}
where $R_{ij}$ is the distance between the top sites, with $i$ and $j$ running over the 5 top sites of each of the capsomers, and $R_{mn}$ is the distance between $m$ and $n$, with $m$ running over the bottom sites of the first capsomer and $n$ running over the top and bottom sites of the second one. $\sigma_\text{t}$ is, as in the original model, the distance between two adjacent top sites in a complete capsid at its lowest energy configuration, and is obtained from the geometry depicted in Fig. \ref{fig:geometry_sigma}:
\begin{equation}
\sigma_\text{t}=2d\sqrt{\frac{1}{10}\left( 5+\sqrt{5}\right)}+2h\sqrt{\frac{1}{10}\left( 5-\sqrt{5}\right)} +R_\text{e}
\label{eq:sigma_value}
\end{equation}
where $d=l\text{sin}(3\pi/10)$.
Similarly, $\sigma_\text{b}$ was initially set to the distance between the top and bottom sites of two adjacent capsomers in a complete capsid, but then was adjusted to $\sigma_\text{b}=0.75\sigma_\text{t}$ to optimize assembly behavior.

\begin{figure}
\begin{center}
\includegraphics[viewport=0 0 605 327, width=0.5\textwidth]{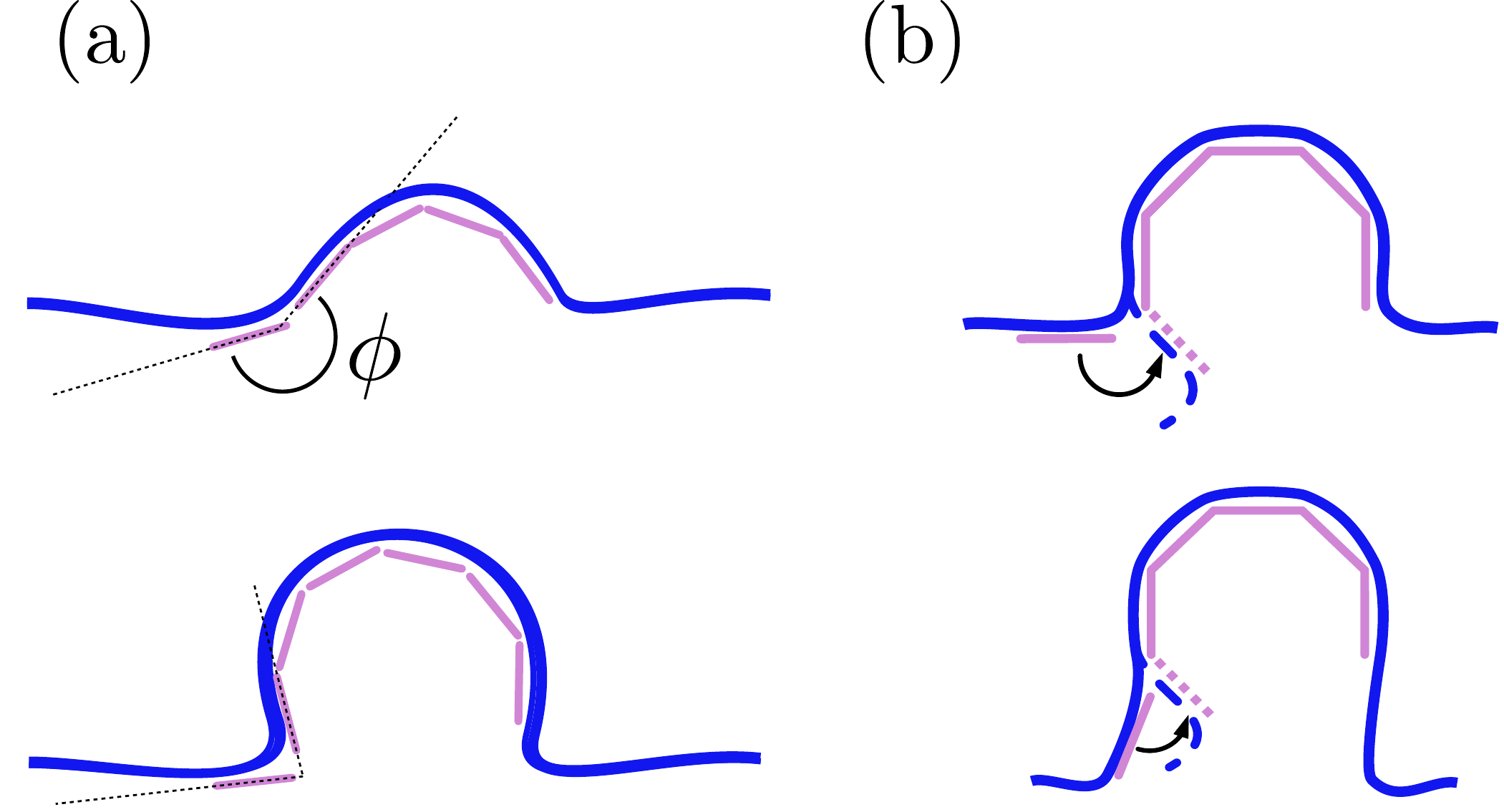}
% geom_assembly.pdf: 605x327 pixel, 72dpi, 21.34x11.54 cm, bb=0 0 605 327

\end{center}
\caption{Geometry of the membrane during simultaneous assembly and budding. {\bf {(a)}} As budding proceeds, the angle between the growing capsid and the associating capsomers becomes more acute. In the original model the long-range repulsions between top beads do not allow sufficient orientational flexibility for capsomers to approach at such acute angles. {\bf {(b)}} Association of a subunit adsorbed on the membrane requires either attachment of the subunit or a local membrane conformational change. {\bf (top)} A short neck formed around a large partial-capsid intermediate leads to a strong kink and thus association of another subunit requires a significant membrane deformation. {\bf (bottom)} A long neck, such as found during assembly of a raft with optimal size, leads to a soft kink and subunit association involves relatively modest membrane deformations.}
\label{fig:geom_assembly}
\end{figure}

\begin{figure}
\begin{center}
\includegraphics[viewport=0 0 696 288, width=0.6\textwidth]{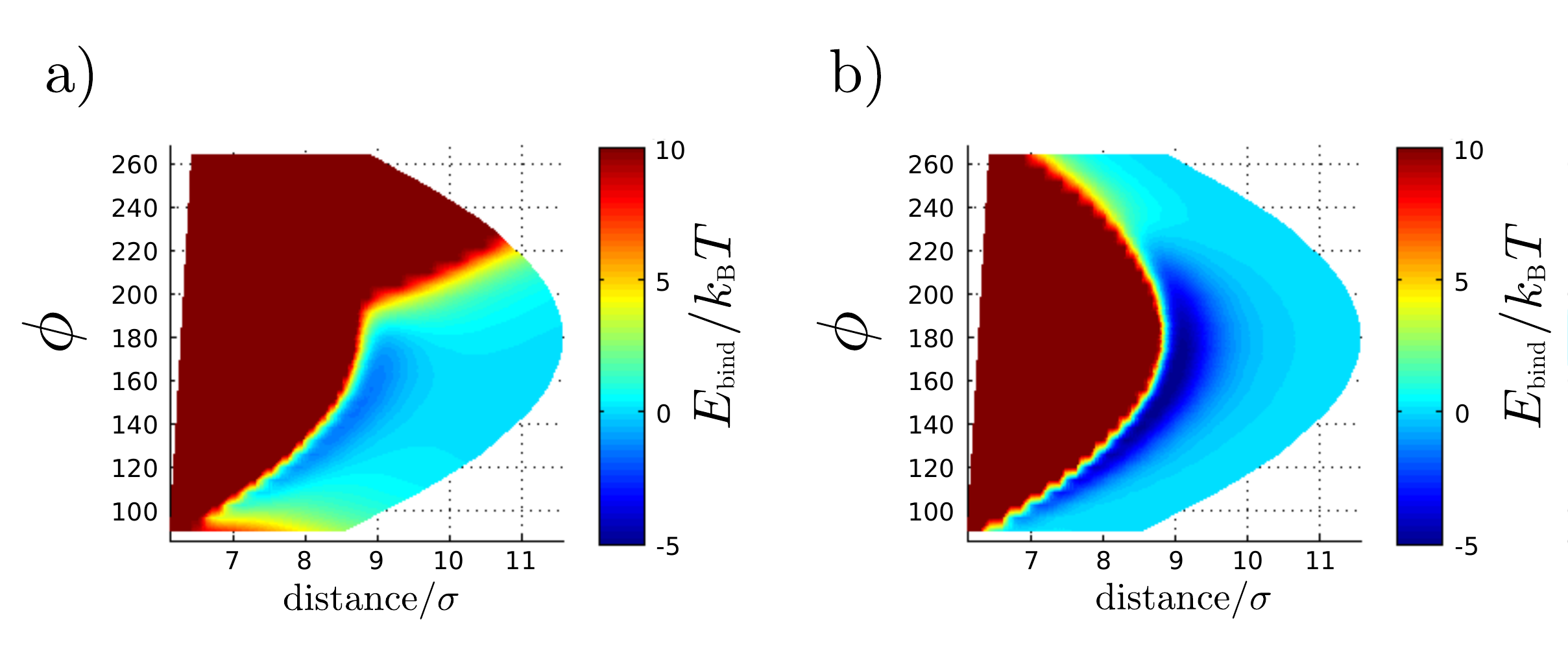}
 % potwales.pdf: 696x288 pixel, 72dpi, 24.55x10.16 cm, bb=0 0 696 288
\end{center}
\caption{Potential energy between two capsomers associating along one edge for different relative angles for {\bf (a)} the original model and {\bf (b)} the extended model. For the original model approach via the acute angles required by association with the membrane leads to large unfavorable potential energies due to the long-range `top'-`top' interactions, while the extended model with shorter repulsive interaction length scales allows approach via a broader range of angles. }
\label{fig:wales_pot}
\end{figure}

We changed the form of the repulsive sites in the original model (one top and one bottom site) to 5 sites for the following reasons. From exploratory simulations, we found that membrane-subunit interactions significantly constrained relative orientations of nearby adsorbed subunits for physically relevant values of the membrane bending modulus. Therefore, association can proceed only through a relatively narrow range of face-face angles (in comparison to the angles available for association in solution). As the partial capsid grows, the accessible range of angles narrows even further (Fig.~\ref{fig:geom_assembly}a). Increasing the number of repulsive sites and moving them closer to the capsomer plane enables a decrease in the interaction range, which allows a wider range of approach angles (Fig. \ref{fig:wales_pot}) while maintaining the equilibrium angle at the same value as for the original model. Moreover, the reduction of the interaction cutoff reduced computation times by nearly a factor of 3.

\subsection{Subunit-membrane interactions}

The potential between capsomers and lipids accounts for attractive and excluded-volume interactions.

\subsubsection{Adhesion interaction sites}

The attractive subunit-membrane interaction is mediated by six sites, one at each of the five vertices and one at the center of the capsomer. Each site sits at a distance $L_{s}$ from the pentamer plane (Fig. \ref{fig:fullcapsomer}). These new interaction sites interact only with the tails of the lipid molecules; in simulations with a domain the sites interact only with domain lipid tails. The attractor-tail interaction is the same as the tail-tail interaction except that there is no repulsive component, as if the attractors were point-particles with no excluded volume:
\begin{equation}
 U_\text{ad} (r)=\left\{ \begin{array}{ccc}
 -\epsilon &;& r<r_\text{c} \\
-\epsilon \cos^{2} \frac{\pi (r-r_\text{c})}{2\omega_\text{c}} &;&r_\text{c}\leq r \leq r_\text{c}+\omega_\text{c} \\
0 & ; & r>r_\text{c}+\omega_\text{c}
\end{array} \right.
\label{eq:adh_hydrophobicity}
\end{equation}
where $r$ is the distance between a capsomer adhesion site and the tail bead of a lipid.

\subsubsection{Excluder sites}

\begin{figure}
\begin{center}
\includegraphics[viewport=0 0 433 198,width=0.45\textwidth]{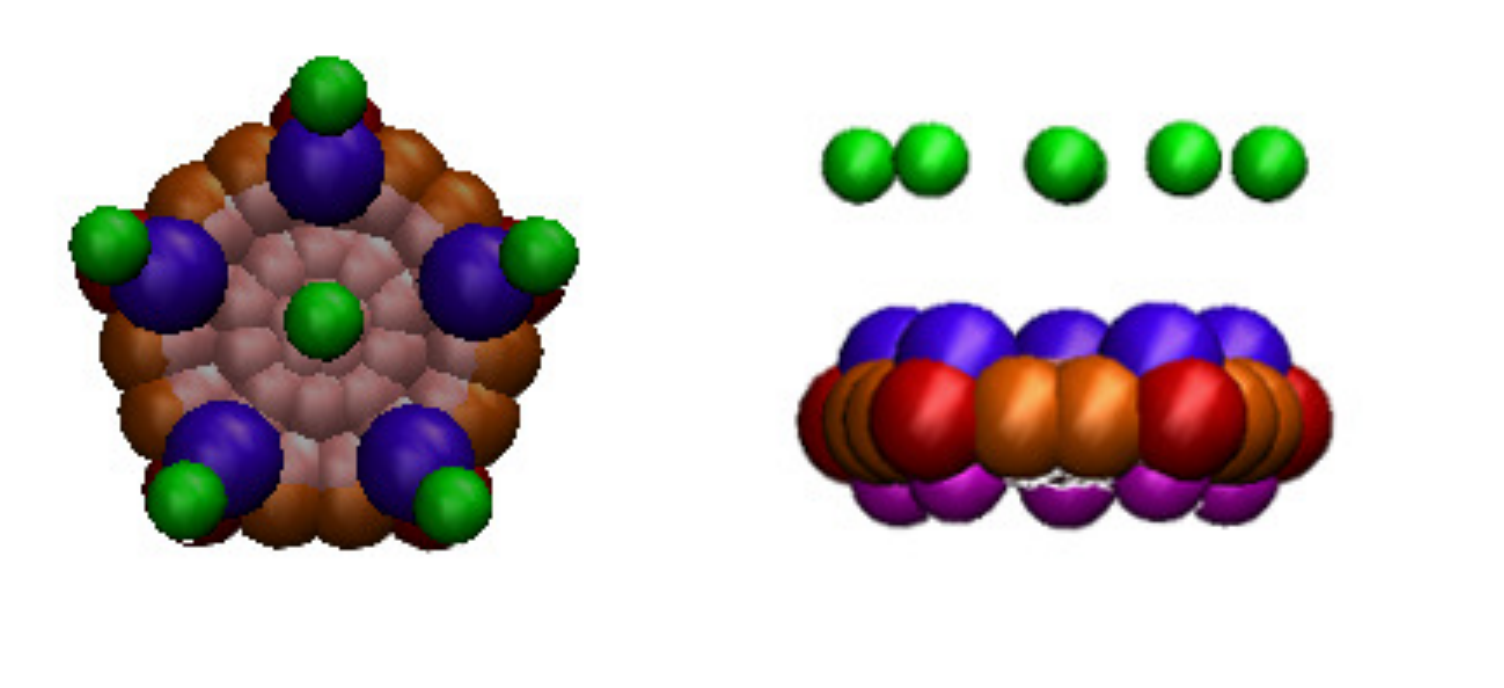}
% capsomerfinal.pdf: 433x198 pixel, 72dpi, 15.28x6.99 cm, bb=0 0 433 198

\end{center}
\caption{Top and side view of the model capsomer, with attractor sites in red and orange, top and bottom repulsive sites in violet and magenta, excluders in pink, and membrane-capsomer interaction sites in green.}
\label{fig:fullcapsomer}
\end{figure}

A layer of 35 beads arranged in the shape of a pentagon is added to the capsomer base to prevent the overlap of viral subunits and membrane lipids (Fig. \ref{fig:fullcapsomer}). These beads interact via an excluded volume potential $U_\text{ex}$ with all lipid beads:

\begin{equation}
 U_\text{ex}(r)=4 \epsilon_{0} \left[ \left( \frac{\sigma}{r-s}\right) ^{12}-\left( \frac{\sigma}{r-s}\right) ^{6}+\frac{1}{4}\right]
 \label{eq:vpt5}.
\end{equation}
with $s=(\sigma_\text{ex}+\sigma)/2-1$, and $\sigma_{ex}$  as the size of the excluders.
In this way, the effective shape of the capsomer is a regular pentagon with thickness $\sigma_\text{ex}$.

\subsubsection{Adhesion energy}
\label{sec:adhesion_energy}

\begin{figure}
\begin{center}
 \includegraphics[viewport=0 0 715 276, width=0.6\textwidth]{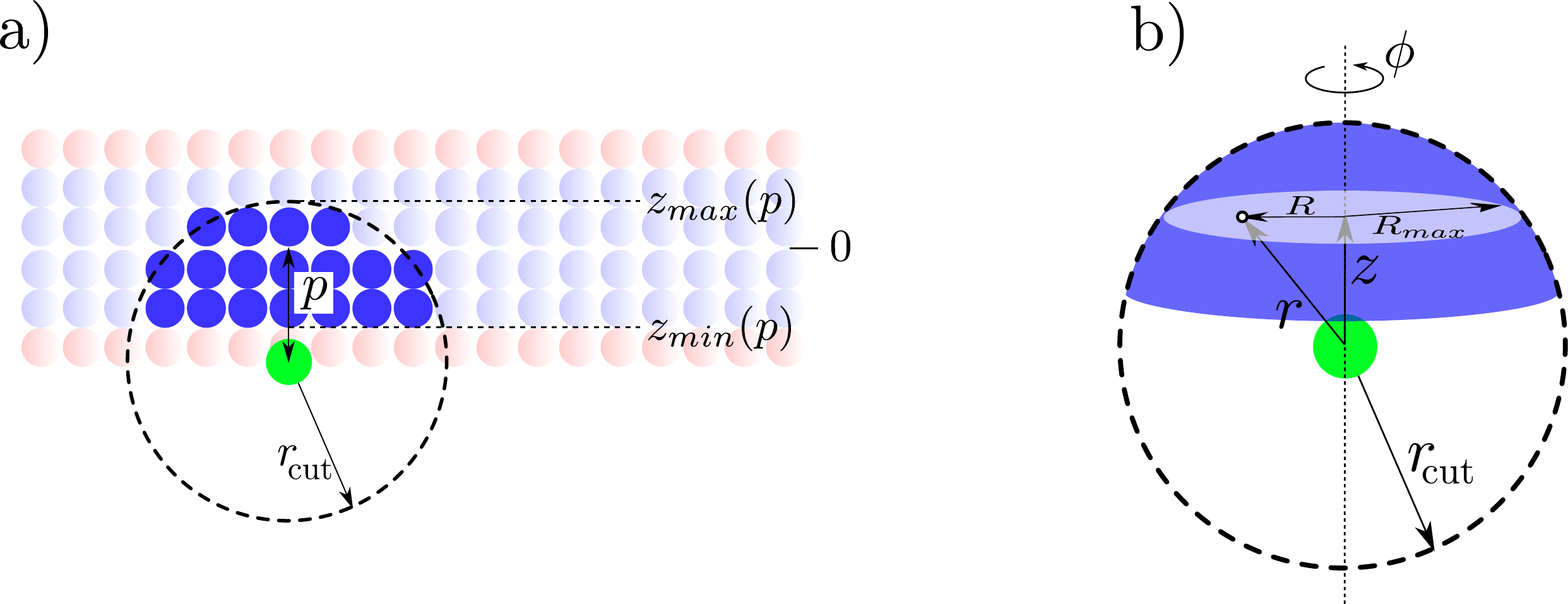}
 % adhesion_energy.pdf: 715x276 pixel, 72dpi, 25.22x9.74 cm, bb=0 0 715 276
\caption{Geometry used for the calculation of the adhesion energy. {\bf a)} The schematic shows a slice of a membrane with the interaction point representing the matrix protein in green at a given penetration $p$. The lipid tails are represented in blue and the heads in red. The tail beads that lie inside the matrix protein interaction volume $V_\text{int}$ are showed in solid blue and they are confined between $z_\text{max}$ and $z_\text{min}$ in the z-direction. The potential cutoff is given by $r_\text{cut}=r_c+\omega_c$. {\bf b)} Geometry used for the integration of the adhesion energy . The energy contribution of every point inside a disk of radius $R_\text{max}(z)$ and width dz is integrated inside the interaction volume (represented in blue).}
\label{fig:adhesion_energy_geo}
\end{center}
\end{figure}

%\commentbis{I have changed all the section}
The adhesion free energy per capsomer was estimated from the calculation of the interaction between the matrix protein attractive site and the lipid tail beads lying inside its interaction range. The number of interacting beads depends on the matrix protein penetration into the membrane (Fig.\ref{fig:adhesion_energy_geo}a), so the free energy was integrated over the accessible values of the penetration $p$:

\begin{equation}
F=-\kt\ln\left(\frac{\int_{p_\text{min}}^{p_\text{max}} e^{-E(p)}dp}{v_0^{1/3}}\right)  %    {p_\text{min}}^{p_\text{max}} E_\text{ad}(p)\text{d}p
\label{eq:adhesion_energy}.
\end{equation}

where $v_0=(\Delta p \pi r_\text{cut}^2)$ is the standard volume, $\Delta p$ the range of possible penetrations where capsomer adhesion sites experience attractive interactions with the membrane, and $E(p)$ the interaction energy for a given penetration:

\begin{equation}
 E(p)=6\rho\int_{z_\text{min}(p)}^{z_\text{max}(p)}\text d z \int_0^{R_\text{max}(z)} U(\vec{r}-p)\text 2\pi r d r
 \label{eq:energy_p}
 \end{equation}
with $\rho$ the density of lipid tails and 6 standing for the number of interaction sites per capsomer. The geometry of the system used for the integration is shown in Fig.\ref{fig:adhesion_energy_geo}b

We found that the adhesion free energy is linearly related to $\epsilon_\text{ad}$:
\begin{equation}
e_\text{ad}\approx \alpha\epsilon_\text{ad}
\label{eq:adhesion_energy_final}
\end{equation}
with $\alpha=-135.3$. Note that this estimate overestimates the adsorption free energy, since it does not include entropic penalties suffered by lipid molecules upon subunit adsorption.
\subsection{Parameters}
\label{sec:parameters_ch5}
The parameters for the membrane are chosen from Ref. \cite{Cooke2005b} so that the bilayer is in a fluid state. We set the temperature of our simulations to $k_\text{B}T/\epsilon_{m} =1.1$ and the lipid-lipid interaction range to $\omega_\text{c}=1.5\sigma_m$, both in equation (\ref{eq:hydrophobicity}) and equation (\ref{eq:adh_hydrophobicity}). The bending rigidity for these values is $\kappa=8.25 k_\text{B}T$ and the areal density of lipids $\eta=0.768\sigma^{-2}$ .

The parameters for the virus model were set according to the phase diagrams of the original model \cite{Johnston2010,Fejer2009} and our exploratory simulations of assembly on a membrane. We found that the optimal parameters that allow large assembly yields for a wide range of concentrations for $k_\text{B}T/\epsilon_\text{v}=1$ are: $R_\text{e}=1\sigma_\text{v}$,
$r_\text{v}=5\sigma_\text{v}$,
$\rho=3$,
$h=0.1875r_\text{v}$,
$l=0.25r_\text{v}$,
$\sigma_\text{b}=0.75\sigma_\text{t}$,
$L_s=1.2r_\text{v}$,
$\epsilon_{att}^v=4.0\epsilon_\text{v}$,
$\epsilon_{att}^e=2.0\epsilon_\text{v}$, and
$\epsilon_{rep}=0.09\epsilon_\text{v}$, with
the subindices `m' and `v' referring to the units of the membrane and virus models respectively.
In order to couple both models, we used the membrane units as the fundamental units and rewrote the parameters for capsid assembly as functions of them. The units of energy, length, and time in our simulations were then respectively $\epsilon_{0}=\epsilon_m$, $\sigma=\sigma_m$ and $\tau_{0}$. The values of the capsid parameters were chosen so that the total energy of assembly exceeds the bending energy of wrapping the capsid. In this way, the unit of energy of the assembly model is set to $\epsilon_\text{v}=2.9\epsilon_0$, the length and time parameters are the same of those of the membrane model, $\sigma_\text{v}=\sigma$ and $\tau_\text{v}=\tau_0$.  Since $k_\text{B}T/\epsilon_v<1$, the energy needed for assembly on the membrane is above the optimal energy for bulk assembly.  Finally, the thickness of the capsomer is $\sigma_\text{ex}=1.25\sigma_0$, and the total mass of a capsomer is $m_\text{capsomer}=66 m_0$.

The remaining parameters can be assigned physical values by setting the system to room temperature, $T=300K$, and noting that the typical width of a lipid bilayer is around 5 nm, and the mass of a typical phospholipid is about 660 g/mol. The units of our system can then be assigned as follows: $\sigma=0.9$ nm, $m_{0}=220$ g/mol, $\epsilon_{0}=3.77 \times 10^{-21} \text{J}=227\text{g}\text{\AA{}}^{2}/\text{ps}^{2}\text{mol}$, and $\tau_{0}=\sigma\sqrt{m_{0}/\epsilon}=8.86$ ps.

%\bibliographystyle{plos2009}
%\bibliography{library2}

\begin{thebibliography}{83}
\bibitem[ten()]{tensionNote}
To ensure that this algorithm maintain constant tension in the presence of
  curved membranes during budding, we measured local areal densities. We found
  that, away from subunits, the mean density and magnitude of fluctuations were
  unchanged throughout budding, indicating that the imposed tension remained at
  zero. We did measure small variations in areal density in the vicinity of
  budding particles. These variations could be explained by subunit-lipid
  interactions and curvature, but might also indicate local nonzero tensions.
  However, variations in local forces should be expected in the vicinity of
  adsorbed particles in physical membranes as well.
\end{thebibliography}

\begin{figure}
\begin{center}

 \includegraphics[viewport=0 0 785 316, width=0.8\textwidth]{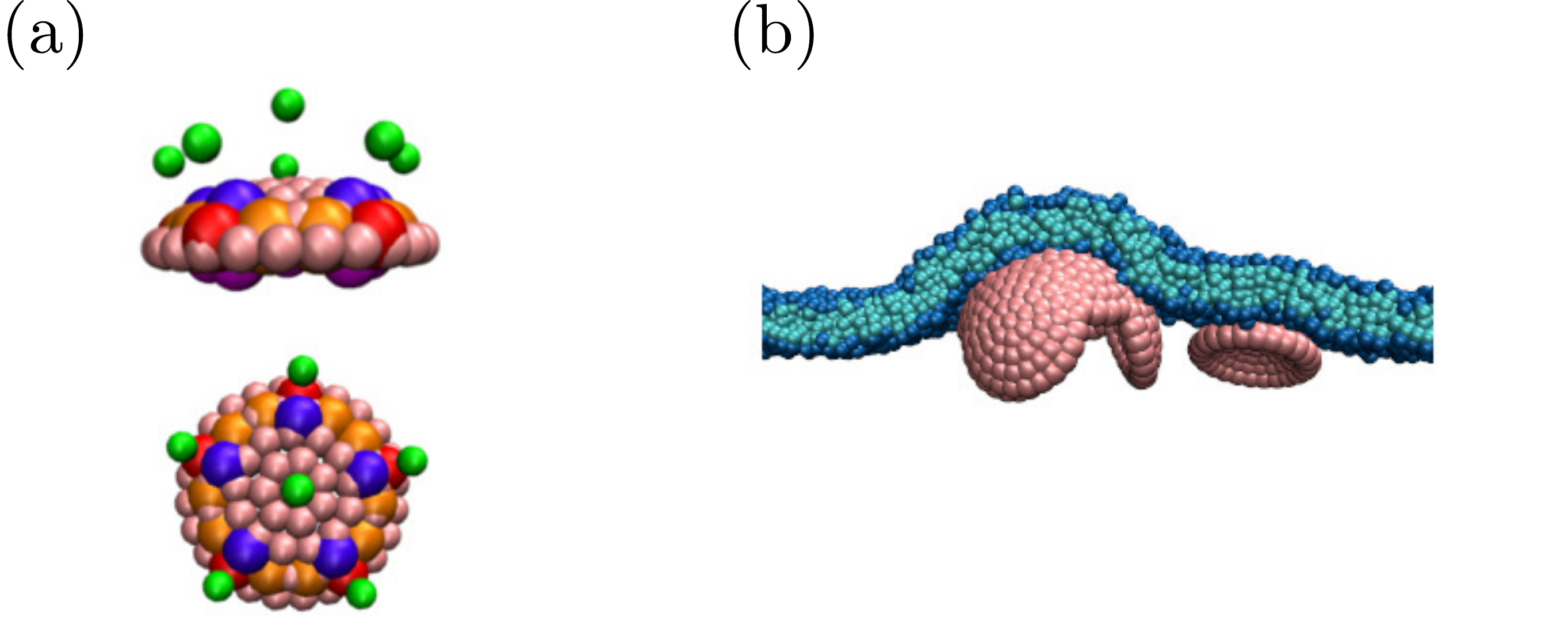}
 % curved_capsomer.pdf: 785x316 pixel, 72dpi, 27.69x11.15 cm, bb=0 0 785 316
\end{center}

\caption{Curved capsomer model. (a) Top and side view of the capsomer. Sites are the same as for the planar subunit; attractive sites are red and orange, top
and bottom repulsive sites are violet and magenta, excluders are pink, and capsomer-lipid interaction sites are green. (b) On a homogenous membrane, assembly stalls at the half capsid, as found for the planar case.
}
\label{fig:curved_capsomer}

\end{figure}

\begin{figure}
\begin{center}
\includegraphics[viewport=0 0 391 225, width=0.5\textwidth]{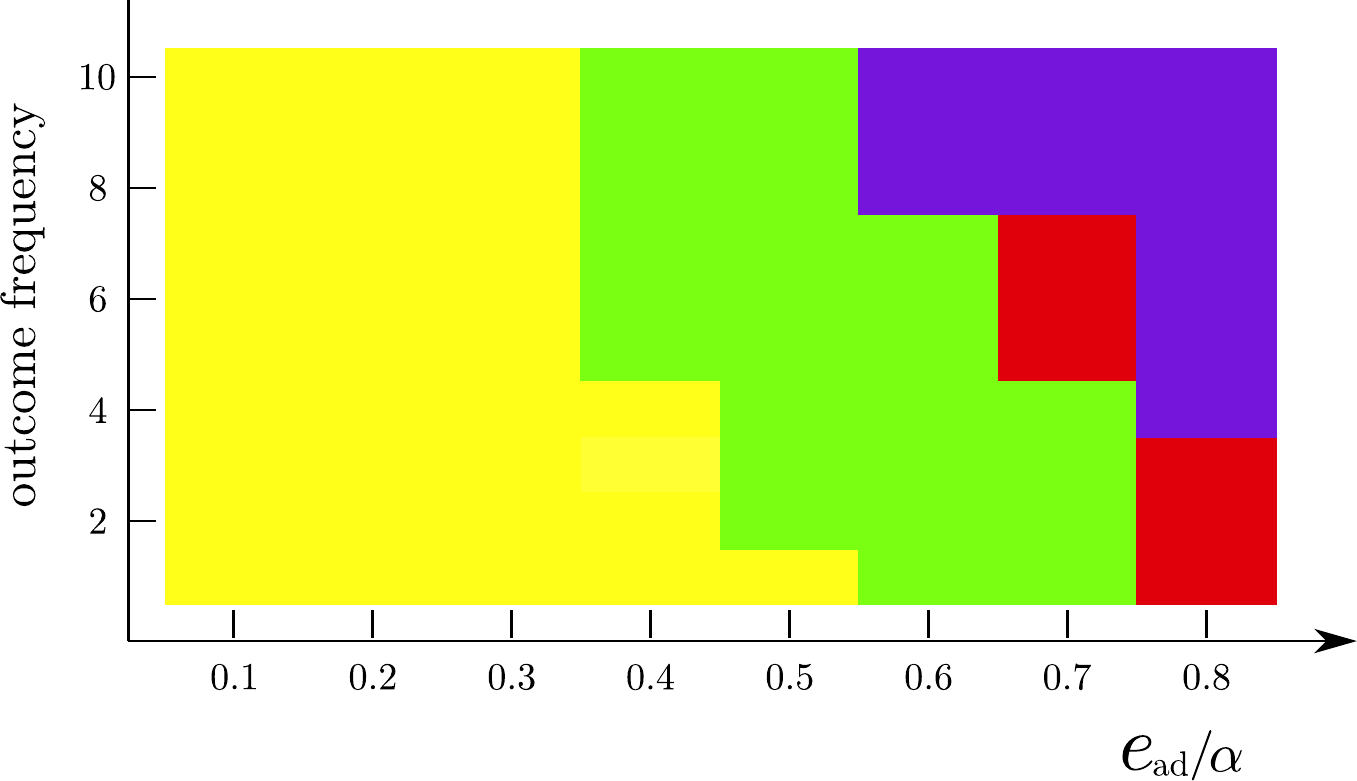}
% cumulative.pdf: 391x225 pixel, 72dpi, 13.79x7.94 cm, bb=0 0 391 225
\end{center}
\caption{ Cumulative histogram of final configurations as a function of the  adhesion strength for a domain with $r_\text{domain}=35\sigma$ and $\gamma=0.15\gamma_0$. The color code represents the outcome type and follows the same format as in Fig.~\ref{fig:phase_diag1} of the main text: succesful assemby (green), budding of a partial capsid (yellow), stalled assemby with wrapping (red) and malformed assembly (violet).}
\label{fig:cumulative}
\end{figure}

\begin{figure}
\begin{center}
\includegraphics[viewport=0 0 894 737, width=0.45\textwidth]{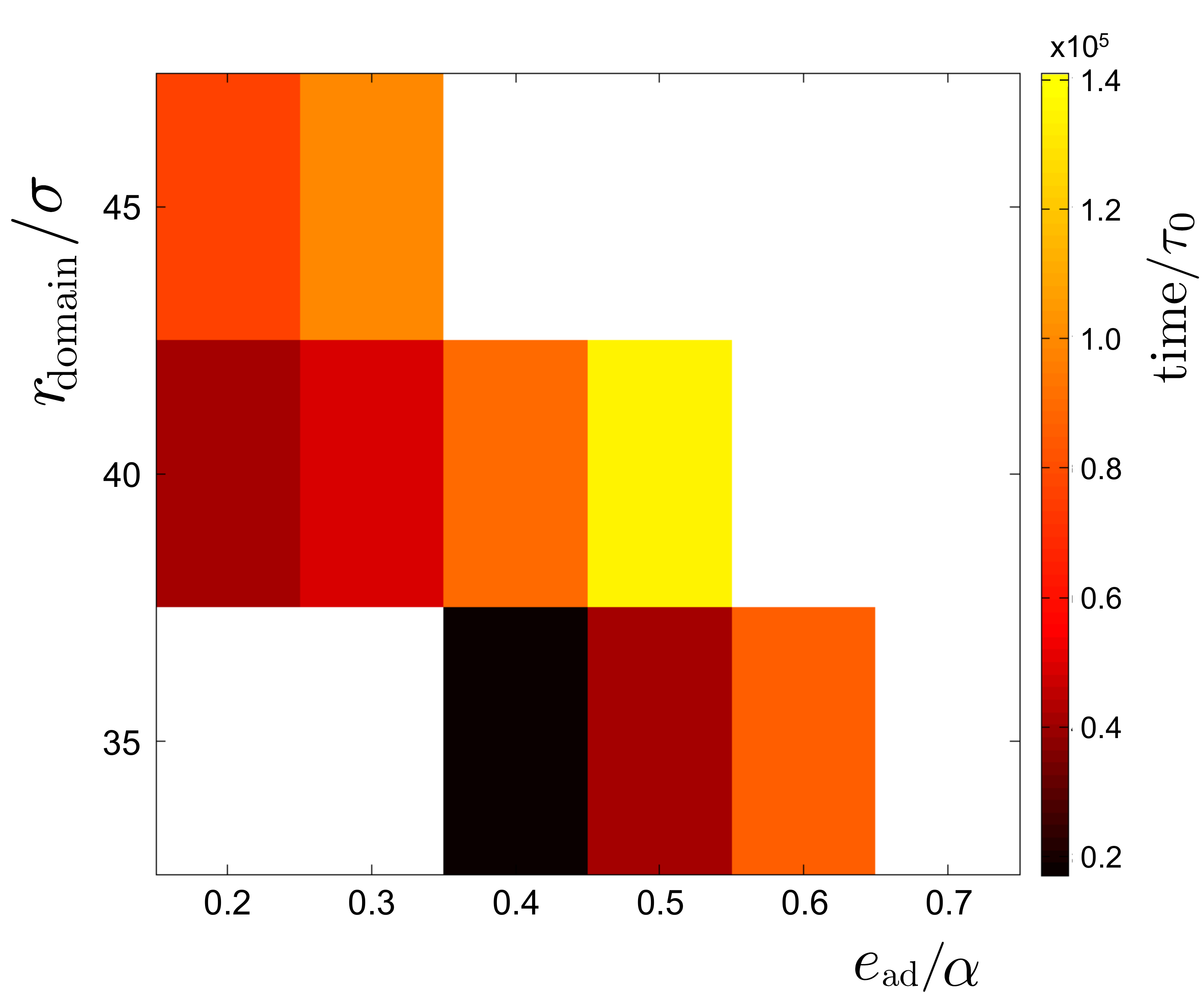}
% times.pdf: 894x737 pixel, 72dpi, 31.54x26.00 cm, bb=0 0 894 737
\end{center}
\caption{Assembly times. The average time needed for an initial assemblage of three capsomers to complete assembly is shown as a function of the domain radius $r_\text{domain}$ and the adhesion energy $\ead$ for $\gamma=0.15\gamma_0$. Grid points shown in white indicate parameter values for which assembly was not completed. }
\label{fig:assembly_time}
\end{figure}

\begin{figure}
\begin{center}
\includegraphics[viewport=0 0 1138 1215, width=0.6\textwidth]{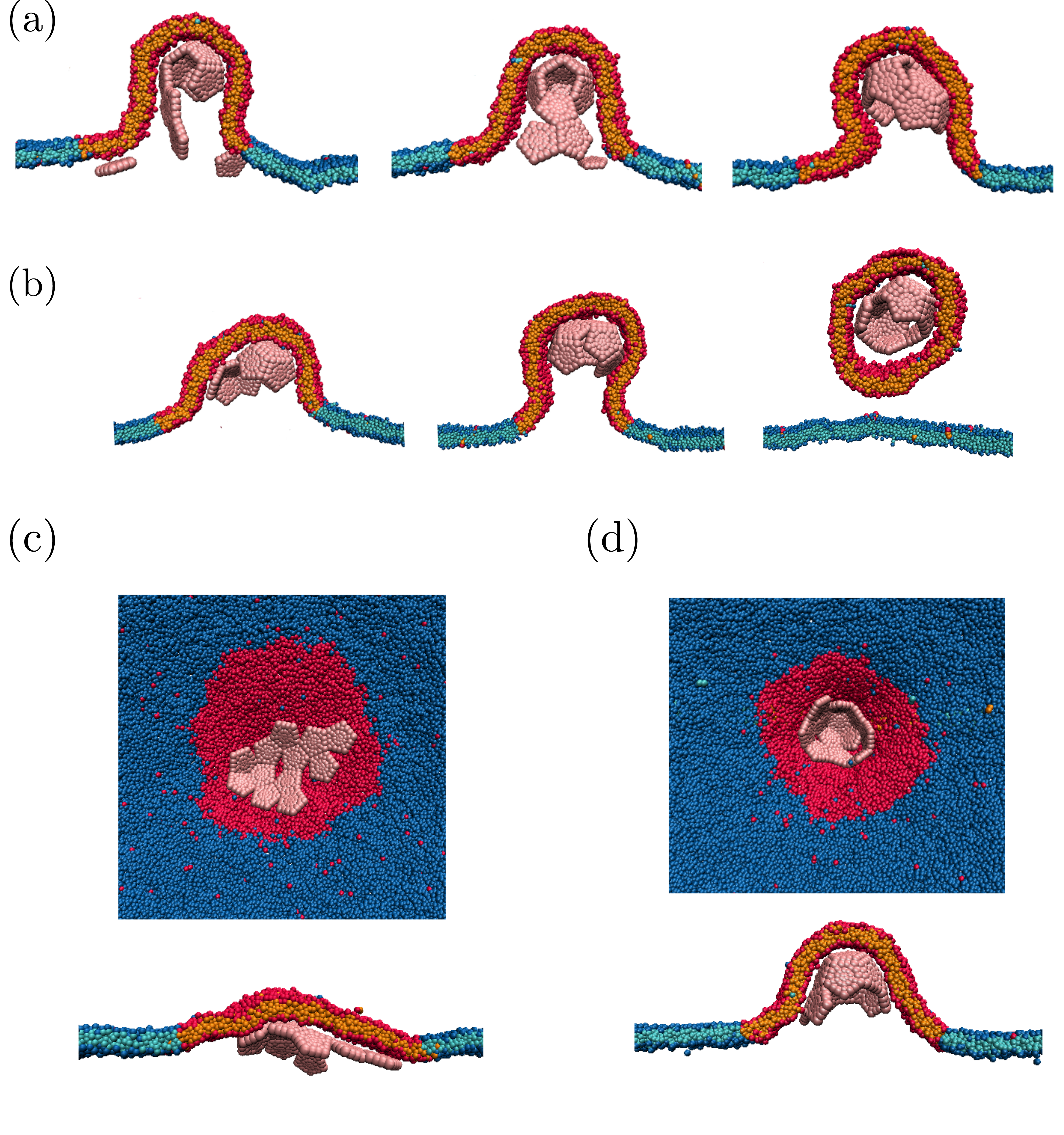}
% kinetic_traps.pdf: 1138x1215 pixel, 72dpi, 40.15x42.86 cm, bb=0 0 1138 1215

\end{center}
\caption{ {\bf Kinetic traps.} Simulation snapshots illustrating some typical kinetic traps, for $r_\text{domain}{=}35\sigma$ and $\gamma{=}0.15\gamma_0$ with varying $\ead$ and time steps between subunit injections $\tinject$. {\bf (a)} Slices of configurations at different times  for $\ead{=}0.5\alpha$ and $\tinject=1500\tau_0$. A dimer associates with a strained geometry to the growing capsid; therefore, the next subunit is prevented from proper association and a malformed capsid arises. {\bf (b)} Two partial capsids nucleate and then coalesce into a malformed assemblage, which then drives budding of the entire domain.  Parameters are $\ead{=}0.3\alpha$ and $\tinject{=}0\tau_0$.   {\bf (c)}  High values for the adhesion strength $\ead{=}0.6\alpha$ and injection rate $\tinject{=}0\tau_0$ lead to formation of a flat aggregate on the membrane. {\bf (d)} An intermediate adhesion strength  $\ead{=}0.5\alpha$ and high injection rate $\tinject{=}0$ lead to formation of a partial capsid trapped within a flat
aggregate. Both top and side views are shown for (c) and (d). }
\label{fig:kinetic_traps}
\end{figure}

\begin{figure}
\begin{center}
\includegraphics[viewport=0 0 1221 601, width=0.75\textwidth]{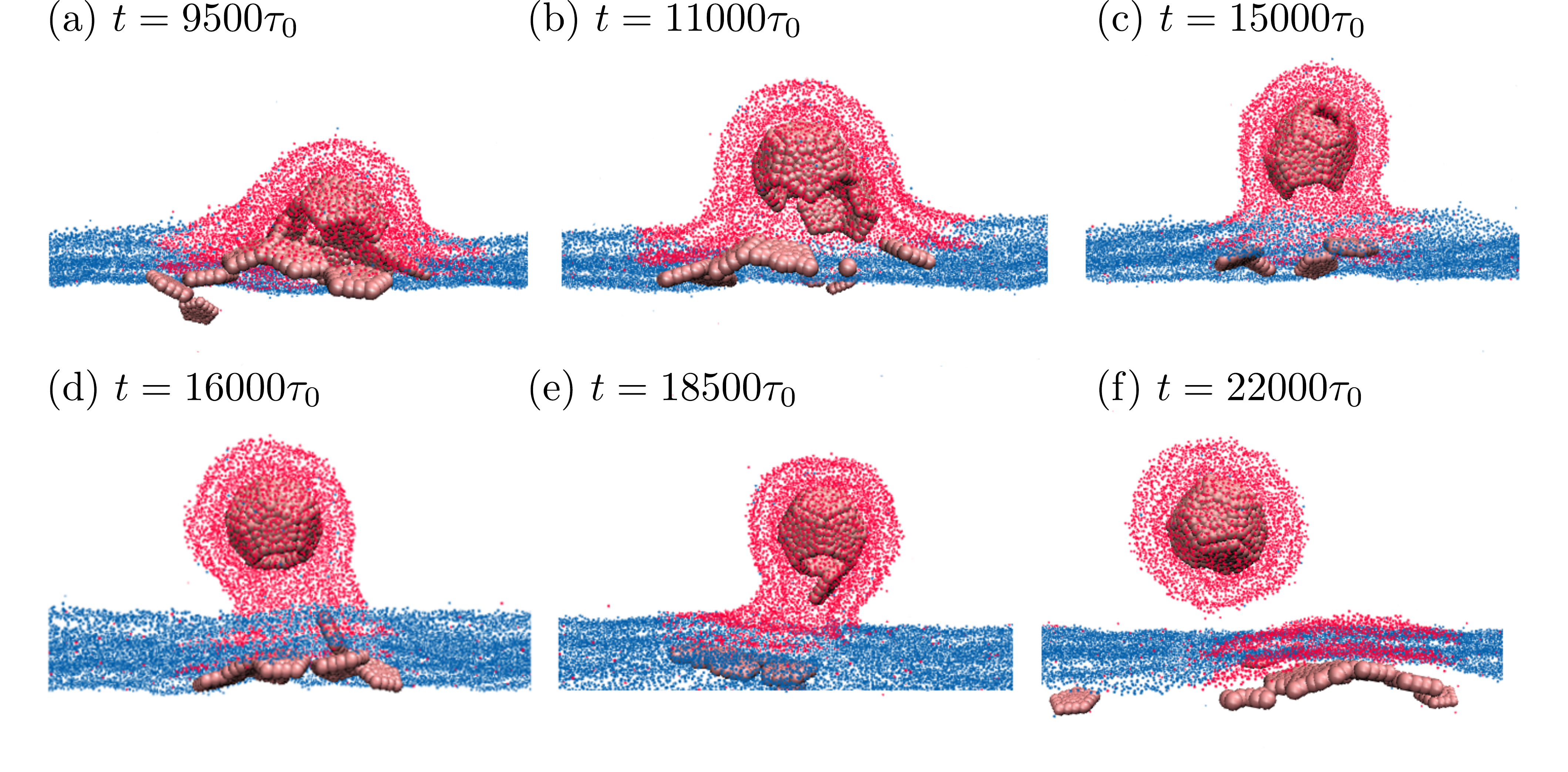}
% bud19.pdf: 1221x601 pixel, 72dpi, 43.07x21.20 cm, bb=0 0 1221 601
\end{center}
\caption{Assembly and budding of a complete capsid in a system with 19 capsomers for $\ead{=}0.3\alpha$ and $\tinject{=}500\tau_0$, $\gamma{=}0.15\gamma_0$, and $\rdomain{=}35\sigma$. Side views of the process are shown, with indicated times since the simulation started ($t=0$).   {\bf (a)} When the last subunit is injected ($t{=}9500\tau_0$), the capsid is already half formed. {\bf (b)} Two partial aggregates are formed, and  {\bf (c)} assemble into a malformed capsid.  {\bf (d)} The capsomers rearrange into an almost finished capsid.  {\bf (e)} The last subunit assembles and  {\bf (f)} the capsid buds }
\label{fig:budplenty}
\end{figure}

\begin{figure}
\begin{center}
\includegraphics[viewport=0 0 374 308, width=0.5\textwidth]{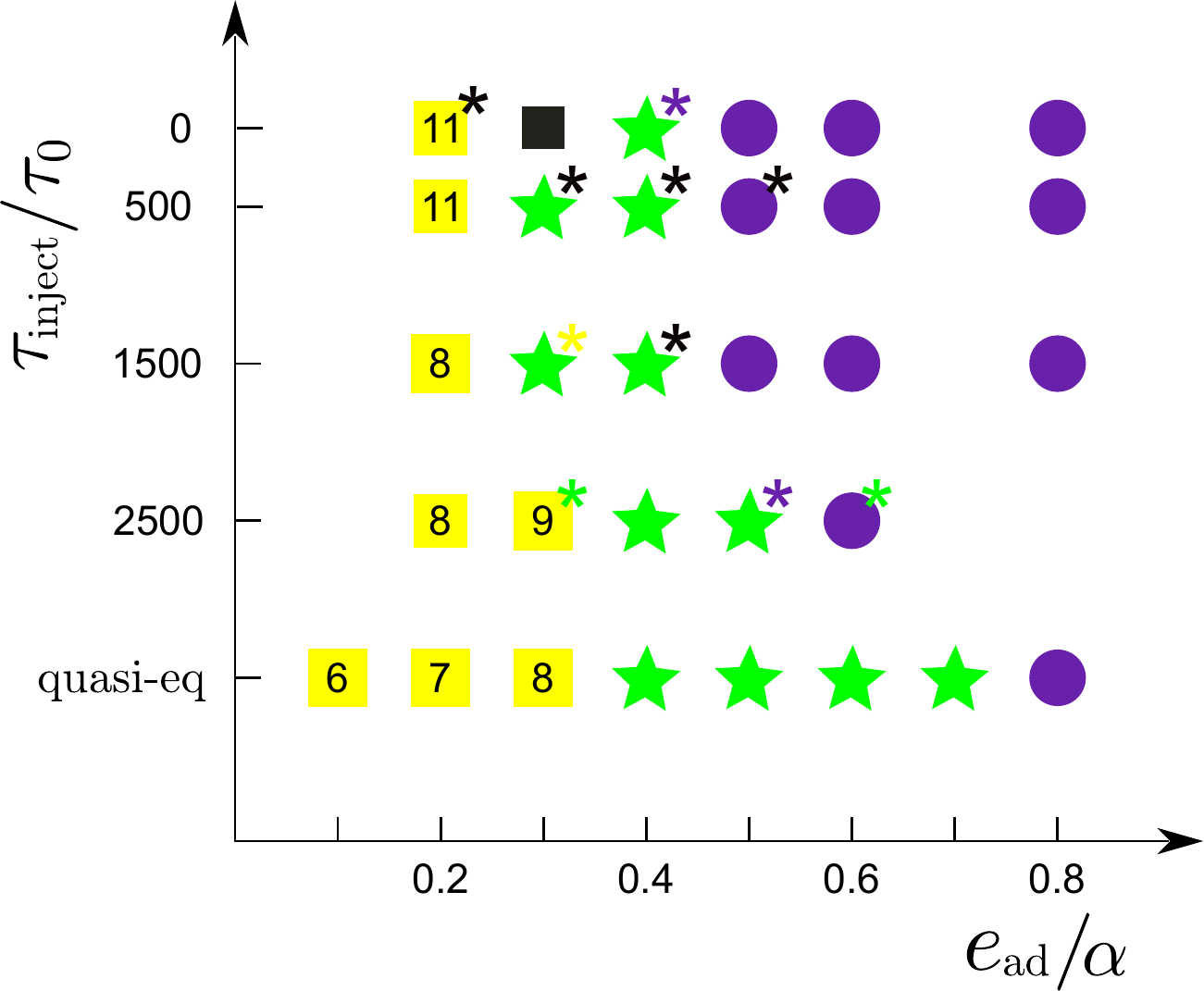}
% phasediagdynamics_soft_asterisks.pdf: 374x308 pixel, 72dpi, 13.19x10.87 cm, bb=0 0 374 308
\end{center}
\caption{ Predominant end products as a function of the subunit injection rate and the adhesion strength for a domain with $r_\text{domain}=35\sigma$ and $\gamma=0.15\gamma_0$.  The most frequent outcome is shown for every set of parameters (with symbols as defined in Fig. \ref{fig:phase_diag1} of the main text). The asterisks indicate that other behaviors are observed in some trajectories, with the asterisk color representing the nature of the alternative outcomes. Red asterisks indicate that some trajectories resulted in incomplete assembly with wrapping, green asterisks indicate that some trajectories resulted in complete assembly and wrapping, and black asterisks indicate that the alternative behavior is the budding of the whole raft with a malformed capsid as shown in Figure \ref{fig:kinetic_traps}b. }
\label{fig:phase_diagdyn_asterisks}
\end{figure}

\begin{figure}
\begin{center}
\end{center}
\caption{The attached movie files show an animation from an assembly trajectory for $\ead{=}0.4\alpha$, $r_\text{domain}=35\sigma$  and $\gamma=0.125\gamma_0$. }
\label{fig:movie}
\end{figure}

\begin{figure}
\begin{center}
\includegraphics[viewport=0 0 1246 448, width=0.5\textwidth]{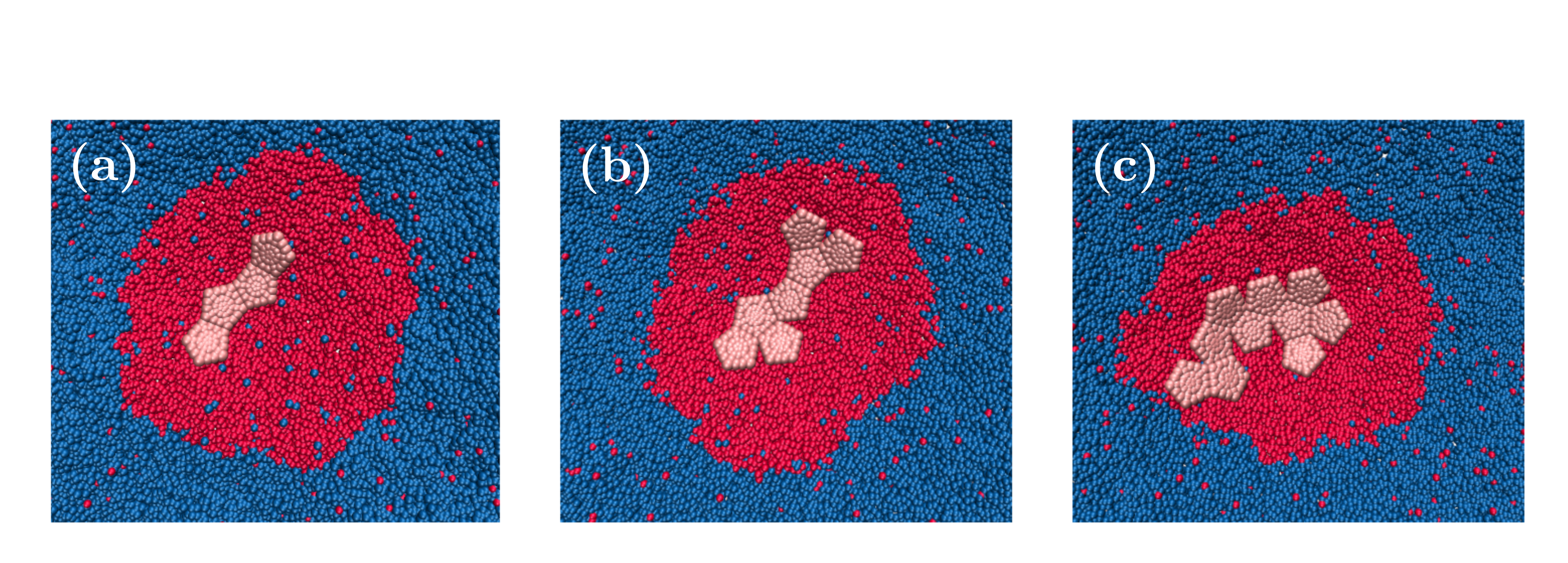}
% flat_line.pdf: 1246x448 pixel, 72dpi, 43.96x15.80 cm, bb=0 0 1246 448
\end{center}
\caption{Typical mechanism for formation of flat aggregates during quasi-equilibrium simulations. For high adhesion strengths and low line tensions, membrane deformation is not promoted, and the capsomers tend to aggregate in a line. This line does not serve as a proper scaffold for budding, the following subbunits associate disorderly and a flat aggregate is formed. Slices of configurations at different times  for $\ead{=}0.6\alpha$ and $\gamma=0.125\gamma_0$ }
\label{fig:assembly_time}
\end{figure}

\end{document}